\documentclass[11pt,a4paper]{article}
\usepackage[utf8]{inputenc}
\usepackage[left=3cm,right=3cm,top=3cm,bottom=3cm]{geometry}

\usepackage{amsfonts}

\usepackage{amsmath}
\usepackage{amssymb}
\usepackage{color}
\usepackage{cite}
\usepackage{graphicx}
\usepackage{subcaption}
\usepackage{slashed}
\allowdisplaybreaks

\usepackage{color}
\definecolor{nicered}{rgb}{0.7,0.1,0.1}
\definecolor{nicegreen}{rgb}{0.1,0.5,.1}

\usepackage[colorlinks=true
,urlcolor=magenta
,anchorcolor=black
,citecolor=blue
,filecolor=black
,linkcolor=red
,menucolor=black
,linktocpage=true
,pdfproducer=medialab
]{hyperref}

\def\Xint#1{\mathchoice
{\XXint\displaystyle\textstyle{#1}}%
{\XXint\textstyle\scriptstyle{#1}}%
{\XXint\scriptstyle\scriptscriptstyle{#1}}%
{\XXint\scriptscriptstyle\scriptscriptstyle{#1}}%
\!\int}
\def\XXint#1#2#3{{\setbox0=\hbox{$#1{#2#3}{\int}$ }
\vcenter{\hbox{$#2#3$ }}\kern-.6\wd0}}

\def\dashint{\Xint-}

\usepackage{stackrel}

\usepackage{rotating}

\usepackage{multirow}

\usepackage{float}

\usepackage{xcolor}

\usepackage{authblk}

\title{Final-state interactions in the CP asymmetries of charm-meson two-body decays}

\date{}

\author[a]{Antonio~Pich}
\author[a]{Eleftheria~Solomonidi}
\author[a]{Luiz~Vale~Silva}

\affil[a]{Departament de F\'{i}sica Te\`{o}rica, Instituto de F\'{i}sica Corpuscular,

Universitat de Val\`encia -- Consejo Superior de Investigaciones Cient\'{i}ficas,

Parc Cient\'{i}fic, Catedr\'{a}tico Jos\'{e} Beltr\'{a}n 2, E-46980 Paterna, Valencia, Spain}

\begin{document}

\hfill IFIC/23-17

{\let\newpage\relax\maketitle}

\maketitle

\textbf{Abstract.}
Urgent theoretical progress is needed in order to provide an estimate in the Standard Model of the recent measurement by LHCb of direct CP violation in charm-meson two-body decays.
Rescattering effects must be taken into account for
a meaningful theoretical description of the amplitudes involved in such category of observables, as signaled by the presence of large strong phases. We discuss the computation of the latter effects based on a two-channel coupled dispersion relation, which exploits isospin-zero phase-shifts and inelasticity parameterizations of data coming from the rescattering processes $ \pi \pi \to \pi \pi $, $ \pi K \to \pi K $, and $ \pi \pi \to K \overline{K} $.
The determination of the subtraction constants of the dispersive integrals relies on the leading contributions to the transition amplitudes from the $ 1/N_C $ counting, where $N_C$ is the number of QCD colours.
Furthermore, we use the measured values of the branching ratios to help in selecting the non-perturbative inputs in the isospin limit, from which we predict values for the CP asymmetries.
We find that the predicted level of CP violation is much below the experimental value.


\section{Introduction}


Symmetries, whether exact or not, played a central role in the formulation of the Standard Model (SM), and offer an avenue to move beyond it.
The violation of Charge-Parity (CP) symmetry in the SM emerges from a single parameter, encoded in the Cabibbo-Kobayashi-Maskawa (CKM) matrix.
Whatever the physics that lies Beyond the SM (BSM) is, it generally introduces new sources of CP violation, challenging the minimal picture depicted by the SM.
Therefore,
a prominent way to hunt for BSM physics consists of studying transitions that change quark flavour, and in particular cases that are sensitive to CP violation.
Being a manifestation of the weak sector of the SM, CP-violating observables are sensitive to high energies, helping to collect hints of BSM dynamics beyond the electroweak scale.

The single CP-violating phase of the Kobayashi-Maskawa (KM) mechanism of the SM must be responsible for CP violation across different flavour sectors.
This mechanism has been tested in the bottom and strange sectors (see Ref.~\cite{Charles:2015gya}), but tests in the charm sector are still missing.
Other than providing novel tests of the KM mechanism,
charm constitutes physics of the up-type and is then complementary to the down-type sector, which is comparatively better known.
In particular, the charm sector offers the opportunity to understand QCD at intermediate energy regimes, namely, in between the light flavours and the bottom, in both of which cases there exist consolidated theoretical tools.
Moreover, with charm physics one can also access flavour-changing neutral currents (FCNCs) of the up-type, where a more effective Glashow-Iliopoulos-Maiani (GIM) mechanism 
applies, 
which represents an opportunity for clear identification of BSM contributions.

In regard of tests of the KM mechanism, CP violation in the charm sector has been established recently by LHCb \cite{LHCb:2019hro}, which measured the difference
of direct CP asymmetries in $D^0$ decays
\begin{equation}\label{eq:CP_discovery_in_charm}
    \Delta A_{CP}^{\mathrm{dir}} = ( -15.7 \pm 2.9 ) \times 10^{-4}
\end{equation}
between final states involving two charged kaons $ A_{CP} (D^0 \to K^- K^+) $,
or two charged pions $ A_{CP} (D^0 \to \pi^- \pi^+) $, where\footnote{Time-integration is left implicit; the contribution of indirect CP violation is negligible \cite{LHCb:2019hro}.}
%
\begin{equation}\label{eq:ACP}
	A_{CP} (i \to f) \equiv \frac{| \langle f | T | i \rangle |^2 - | \langle \overline{f} | T | \overline{i} \rangle |^2}{| \langle f | T | i \rangle |^2 + | \langle \overline{f} | T | \overline{i} \rangle |^2} = \Sigma_j \left[ p_j \, \sin (\Delta \delta_j) \, \sin (\Delta \phi_j) \right]_{i \to f} \,,
\end{equation}
$T$ being the transition matrix. 
In order to have a non-vanishing CP asymmetry, one needs both differences of weak ($ \Delta \phi $) and strong ($ \Delta \delta $) phases, as indicated schematically in the right-hand side of Eq.~\eqref{eq:ACP}; therein, the sum consists of all possible interference terms $j$ among pairs of amplitudes that have simultaneously different weak ($ \Delta \phi_j \neq 0 \; (\text{mod} \, \pi) $) and strong phases ($ \Delta \delta_j \neq 0 \; (\text{mod} \, \pi) $), and $p_j$ scales like the ratio of a CP-odd over a CP-even amplitudes. 
Weak phases flip sign under CP transformation, while strong phases are left unchanged.
There is an active experimental research program, as attested by the following very recent results \cite{LHCb:2022vcc}:
\begin{eqnarray}\label{eq:ExpCPasym}
    && A_{CP} ( D^0 \to \pi^- \pi^+) = ( +23.2 \pm 6.1 ) \times 10^{-4} \,, \\
    && A_{CP} ( D^0 \to K^- K^+) = ( +7.7 \pm 5.7 ) \times 10^{-4} \nonumber
\end{eqnarray}
(which are correlated at the level of $ 0.88 $, and based on Eq.~\eqref{eq:CP_discovery_in_charm}).
There are also available bounds on CP violation for many other channels (see App.~\ref{app:num_inputs}).
Much progress is expected in the years to come, thanks to LHCb and Belle II, which will largely improve the sensitivity to sources of CP violation;
also, BESIII has an active research program in charm physics.
As a benchmark, the accuracy in some CP asymmetries will be improved by about one order of magnitude.

On the other hand, theory has to match the observed experimental progress.
As previously stated, in the SM the weak phase comes from the CKM matrix.
It is yet unknown whether the source of CP violation therein can explain the measurement of $ \Delta A_{CP}^{\mathrm{dir}} $, or whether this observable signals the emergence of non-SM sources of CP violation: this is due to the presence of non-perturbative QCD effects that are extremely challenging to describe, precluding precision flavour studies at the present moment.
A dynamical mechanism for the generation of the strong phases is the rescattering of on-shell particles, in particular pion and kaon pairs.
It cannot be stressed enough how important the role played by the
strong phases in describing CP asymmetries is.
Indeed, large strong phases generated in such dynamical way via rescattering effects are also associated to large modulations of the amplitudes, that must therefore be fully taken into account in predictions of the SM amplitudes.
The main interest of this work is the determination of these non-perturbative effects, and their impact on the prediction of the CP asymmetry.


A similar problem happens in the case of kaon decays. The SM description of the 
measured 
direct CP violation therein requires the introduction of non-perturbative QCD inputs.
Such inputs can be determined via the use of Dispersion Relations (DRs) \cite{Pallante:1999qf,Pallante:2000hk}.
The analysis is simpler compared to charm-meson decays, since the only relevant final state accessible from kaon decays are pion pairs, motivating an elastic analysis.
In this case, Watson's theorem \cite{Watson:1954uc} applies, and the DRs have a known explicit analytical solution \cite{Muskhelishvili,Omnes:1958hv}.
Moreover, one also disposes of a well established effective field theory, which is chiral perturbation theory ($\chi$PT) for the three lightest flavours \cite{Gasser:1984gg,Ecker:1994gg,Pich:1995bw}.
In order to ensure the convergence of the dispersive integrals and to limit the dependence on the high-energy domain, DRs are eventually ``subtracted'', and $\chi$PT provides the subtraction constants of DRs.
Alternatively, $\chi$PT provides a framework in which rescattering effects can be computed perturbatively.
It is then apparent that DRs provide the resummation of infrared chiral logarithms, which are process independent, while subtraction constants encode the process-dependent ultraviolet dynamics.
Importantly, both approaches show a good agreement \cite{Pallante:1999qf,Pallante:2000hk,Pallante:2001he,Cirigliano:2003nn,Cirigliano:2003gt,Gisbert:2017vvj,Cirigliano:2019cpi}.



In the case of charm physics, we will also employ DRs, which result from two basic principles of any Quantum Field Theory: analyticity (due to causality) and unitarity.
In the present case, however, the required analysis is non-elastic because 
the $D^0$ mass lies well
above the threshold for production of kaon pairs. We have then a set of integral equations related by unitarity.
These equations have to be solved numerically, as no explicit analytical form of the solution is known in general.
We are going to include in our analysis only pion and kaon pairs, for which we dispose of abundant data, and neglect further channels in this work.
Dealing with other channels requires a different set of techniques, that we 
postpone to future work.
Having pions and kaons, we need as inputs two phase-shifts and one inelasticity, which accounts for the probability of transition between pion and kaon pairs;
we use available parameterizations for them 
\cite{Kaminski:2006qe,GarciaMartin:2011cn,Pelaez:2019eqa,Pelaez:2020gnd}.
As in the elastic case of kaon decays into pion pairs, we also need some physical input for the subtraction constants.
We employ large-$N_C$ counting for their determination, based on an expansion in powers of $1/N_C$ with $N_C$ the number of QCD colours \cite{tHooft:1973alw,tHooft:1974pnl,Witten:1979kh}, which is known to provide an understanding of many observed features of non-perturbative strong dynamics \cite{Manohar:1998xv,Pich:2002xy}.
Preliminary results were communicated in Refs.~\cite{ValeSilva:2022kkv,Solomonidi:2023vue}.

Phase-shifts and inelasticity at the energy $M_D$ have been applied, non-dispersively, in e.g. Refs.~\cite{Bauer:1986bm,Franco:2012ck,Bediaga:2022sxw}.\footnote{Note that Ref.~\cite{Bediaga:2022sxw} writes for isospin-zero:

\begin{equation}
    \begin{pmatrix}
    \mathcal{A}_{D^0 \to \pi \pi} \\
    \mathcal{A}_{D^0 \to K K} \\
    \end{pmatrix}
    = S_S
    \begin{pmatrix}
    V^\ast_{cd} V_{ud} \, a_{\pi \pi} \\
    V^\ast_{cs} V_{us} \, a_{K K} \\
    \end{pmatrix}
\end{equation}
with $a_{\pi \pi}, a_{K K}$ real, which seems not to implement the result expected for the strong phase from Watson's theorem in the limit where the rescattering process is elastic.
Also note that Ref.~\cite{Bauer:1986bm} writes $A = S_S^{1/2} A^{\rm bare}$, where $S_S^{1/2}$ encodes the rescattering part, implementing correctly that limit. For a discussion of the latter approach, see Ref.~\cite{Nierste:2016fnj}.} Although they recognize the importance of rescattering effects, these approaches do not capture their full picture, which is the aim of employing a dispersive treatment.
Previous discussions of DRs in the context of charm-meson decays include Refs.~\cite{Sorensen:1981vu,Reid:1981jr,Kamal:1987nm,Kamal:1990ij,Czarnecki:1991ys}, which have not addressed CP violation, being the main focus here. Compared to these references, we discuss DRs and the inputs that we employ in details.



Various other non-dispersive analyses have also been made for the description of multiple charm-meson decay modes, such as topological approaches, the use of $SU(3)_F$ or its sub-groups, transitions assisted by intermediate resonances, etc.; see
Refs.~\cite{Zenczykowski:1996bk,Zenczykowski:1999cu,Brod:2011re,Grossman:2012ry,Hiller:2012xm,Schacht:2022kuj,Soni:2019xko,Schacht:2021jaz,Buccella:2019kpn,Li:2019hho,Cheng:2019ggx,Muller:2015lua,Muller:2015rna,Nierste:2015zra,Wang:2022nbm}.





Also note that calculations based on QCD light-cone sum rules \cite{Khodjamirian:2017zdu} indicate that the SM cannot account for the large level of CP asymmetry observed by LHCb. However, light-cone sum rules have not been extensively tested in the charm sector, requiring alternative methods to support such an extraordinary claim.



Let us also mention that,
although methods to deal with rescattering in the lattice \cite{Hansen:2012tf} are progressing fast, the typical energy scale of charm processes still represents an overwhelming problem for Lattice QCD methods.

Having stressed the need for dealing with strong interactions, let us point out that
there are ways, however, of extracting properties of weak interactions without the need to describe in details the strong dynamics.
In the charm sector, we are not at that stage yet:
%
%
we cannot rely on a strategy such as, for instance, the one employed in the extraction of the unitarity angle $\alpha$ from charmless $B$-meson (quasi-)two-body decays having pions and rhos in the final state, since we do not dispose of the necessary number of measurements at the required level of accuracy to use an isospin analysis \cite{Grossman:2012eb}.

Conversely,
the problem we deal with here is less a question of precision as it is in the case of bottom physics, for instance.
In that sector, one will face in the (near) future the need for better describing sub-leading effects (e.g., long-distance penguin effects in the extraction of $\beta$, better controlling experimental systematics from decays of charm-mesons in the extraction of $\gamma$, dealing with isospin-breaking in the extraction of $\alpha$, etc.).
Rather, in the charm sector
we cannot rely on the experimental (such as isospin analysis) and theoretical (such as heavy quark expansion, due to the slower convergence of the perturbative series) approaches already employed in the other flavour sectors.
It is our goal to employ a data-driven formalism, embodied by the use of DRs.

To conclude this introduction, note that the large level of CP violation observed in $ \Delta A_{CP}^{\mathrm{dir}} $ has triggered studies of contributions from BSM, see Refs.~\cite{Chala:2019fdb,Dery:2019ysp,Bause:2022jes} for recent studies.

This article is organized as follows: in Sec.~\ref{sec:weak_interactions} we set the relevant weak interactions; in Sec.~\ref{sec:DRs} we introduce the DRs; their necessary inputs are discussed in Sec.~\ref{sec:inputs_DRs}, and the numerical solutions of the DRs are given in Sec.~\ref{sec:sols_coupled_channel_DRs}, while the subtraction constants of once-subtracted DRs are discussed in Sec.~\ref{sec:subtraction_constants}; in Sec.~\ref{sec:physical_predictions} we discuss the available mechanisms of CP violation, and give the predictions for the CP asymmetries; conclusions follow in Sec.~\ref{sec:conclusions}. A series of appendices discuss more technical aspects, and fix possible conventions.

\section{Effective weak interactions}\label{sec:weak_interactions}


The full Hamiltonian at low energies contains (renormalizable) strong and electromagnetic interactions, the kinematic terms for the light quarks and the charm quark (including their masses), and (non-renormalizable) effective weak interactions.
The effective interaction Hamiltonian density for $ \Delta C = 1 $ up to operators of dimension-six, valid for energy scales $ \mu_b > \mu > \mu_c $ ($ \mu_q $ being the energy scale at which the quark of flavour $q$ is integrated out), is the following \cite{Buchalla:1995vs}:\footnote{Indices $ 1 $ and $ 2 $ are exchanged with respect to Ref.~\cite{Buchalla:1995vs}, and $C_{1,2}$ ($C_{3, \ldots, 6}$) are called $z_{2,1}$ (respectively, $v_{3, \ldots, 6}$) therein. We are not including in the effective Hamiltonian of Eq.~\eqref{eq:H_eff} neither electroweak penguins nor the electromagnetic dipole.}
%
%
\begin{equation} \label{eq:H_eff}
	\mathcal{H}_{\rm eff} = \frac{G_F}{\sqrt{2}} \left[ \, \sum^2_{i = 1} C_i (\mu) \left( \lambda_d Q_i^d + \lambda_s Q_i^s \right) - \lambda_b\, \sum^6_{i = 3} C_i (\mu) Q_i + C_{8g} (\mu) Q_{8g} \right] + \mathrm{h.c.}
\end{equation}


\noindent where


\begin{equation}
	\lambda_q = V^\ast_{c q} V_{u q} \,, \qquad q = d, s, b \,.
\end{equation}
Unitarity of the $ 3 \times 3 $ CKM matrix $V$ implies:

\begin{equation}
	\lambda_d + \lambda_s + \lambda_b = 0\, .
\end{equation}

The basis of operators is the following:

\begin{eqnarray}\label{eq:operator_list}
	&& Q_1^d = ( \bar{d} c )_{V-A} ( \bar{u} d )_{V-A} \,, \\
	&& Q_2^d = ( \bar{d}_j c_i )_{V-A} ( \bar{u}_i d_j )_{V-A} \stackrel[]{Fierz}{=} ( \bar{u} c )_{V-A} ( \bar{d} d )_{V-A} \,, \nonumber\\
	&& Q_1^s = (\bar{s} c)_{V-A} (\bar{u} s)_{V-A} \,, \nonumber\\
	&& Q_2^s = (\bar{s}_j c_i)_{V-A} (\bar{u}_i s_j)_{V-A} \stackrel[]{Fierz}{=} (\bar{u} c)_{V-A} (\bar{s} s)_{V-A} \,, \nonumber\\
	&& Q_3 = ( \bar{u} c )_{V-A} \sum_q^{} ( \bar{q} q )_{V-A} \,, \nonumber\\
	&& Q_4 = ( \bar{u}_j c_i )_{V-A} \sum_q^{} ( \bar{q}_i q_j )_{V-A} \stackrel[]{Fierz}{=} \sum_q^{} ( \bar{q} c )_{V-A} ( \bar{u} q )_{V-A} \,, \nonumber\\
	&& Q_5 = ( \bar{u} c )_{V-A} \sum_q^{} ( \bar{q} q )_{V+A} \,, \nonumber\\
	&& Q_6 = ( \bar{u}_j c_i )_{V-A} \sum_q^{} ( \bar{q}_i q_j )_{V+A} \stackrel[]{Fierz}{=} -2 \, \sum_q^{} ( \bar{q} c )_{S-P} ( \bar{u} q )_{S+P} \,, \nonumber\\
	&& Q_{8g} = - \frac{g_s}{8 \pi^2} m_c \bar{u} \sigma_{\mu \nu} (1 + \gamma_5) G^{\mu \nu} c \,, \nonumber
\end{eqnarray}
where $ (V \pm A)_\mu = \gamma_\mu (\mathbf{1} \pm \gamma_5) $, $ S \pm P = \mathbf{1} \pm \gamma_5 $, and $i, j$ are colour indices.
The SM Wilson coefficients are fully known to next-to-leading order (NLO)
in perturbative QCD, with some NNLO ingredients available \cite{deBoer:2016dcg}.
Their values are given in App.~\ref{app:num_inputs}.\footnote{We indicate Fierz rearrangements when introducing the basis of operators for later convenience; the Wilson coefficients are calculated at the NLO in the un-Fierzed basis. The gluonic dipole does not affect the Wilson coefficients of the penguin operators at NLO in perturbative QCD}.
Due to the GIM mechanism, 
(short-distance) penguin operators are absent at scales $ \mu > \mu_b $, and result from the NLO matching at $\mu_b$, and the running from $ \mu_b $ to $ \mu_c $; they come with small Wilson coefficients and thus give suppressed contributions to CP violation.
When rescattering effects are large,
the main contribution to the CP asymmetries is expected to come from the non-unitarity of the $ 2 \times 2 $ CKM sub-matrix, see Sec.~\ref{sec:mechanism_CPV} below.
This should be contrasted to the case of bottom physics, where rescattering effects are comparatively much smaller, possibly allowing for perturbative treatments.

\section{Dispersion Relations}\label{sec:DRs}


In describing $D \to \pi \pi, K \overline{K}$
to first order in weak interactions
a discontinuity equation can be written for the transition amplitudes analytically extended to the complex plane (of the invariant mass squared $s$ of the pseudoscalar pair).
The discontinuity is set by the rescattering of the light particles that are stable under strong interactions, with the right-hand cut starting at the threshold for production of pion pairs, and no left-hand cut for the transition amplitudes; for an introduction, see Ref.~\cite{Oller:2019rej}.
The strong dynamics is non-perturbative in nature and has some useful properties: it conserves flavour, C, P, CP, isospin
and G-parity.
The rescattering among light, stable final states gives origin to the strong phases necessary for a non-vanishing CP asymmetry.
In the elastic limit, such a phase in the weak decay can be extracted directly from the phase-shift in the rescattering of pions.
More can be learnt about the rescattering by exploiting its analyticity in the relevant kinematical variables, relating the dispersive/real and absorptive/imaginary parts of the rescattering amplitudes.


In the little Hilbert space (once the global energy-momentum conservation condition has been factored out), the total $S$ matrix can be written as $S={\bf 1} + i\, T$, which implies the unitarity relation 
$T-T^\dagger = i\, T T^\dagger = i\, T^\dagger T$.
In our particular case, $S$ and $T$ are $3\times 3$ matrices describing all possible transitions among the basis of initial and final states $\{ D, \pi\pi , K \overline{K}\}$. 
Restricting to the $\{\pi\pi , K \overline{K}\}$ subspace, the partial-wave (and isospin) projected strong $S_S$ matrix can be written in the form:
%
\begin{equation}
(S_S)^I_J\, =\,  ({\bf 1} + i\, T_S)^I_J \, =\, {\bf 1} + 2 i \;\Sigma^{1/2} (s)\; T^I_J(s)\; \Sigma^{1/2} (s)\, .
\end{equation}
$ S_S $ satisfies the unitarity relation $ S_S^\dagger S_S = S_S S_S^\dagger = {\bf 1} $, 
and $ T_S $ inherits $ T_S - T^\dagger_S = i T_S T^\dagger_S = i T^\dagger_S T_S $.
Since the decaying $D$ mesons are spinless, the total angular momentum of the daughter pair of pseudoscalars is $J=0$. Owing to Bose symmetry, the two-pion state can have isospin $I=0$ and $2$; the isospin of the kaon pair can take the values $I=0, 1$.
Thus, there are two different isosinglet states that get coupled through the rescattering dynamics.
The kinematic factors $\Sigma_i(s) = \Theta (s - 4 M^2_i) \,\sigma_i (s)$ 
incorporate the threshold conditions and the mass corrections to the two-body center-of-mass three-momenta. In the two-channel isosinglet ($I=0$) case, $\Sigma (s)$ becomes a $2\times 2$ matrix:
\begin{equation}
	\Sigma (s) = {\rm diag} \!\left[ \Theta (s - 4 M^2_\pi)\,\sigma_\pi (s)\, , \,\Theta (s - 4 M^2_K)\,\sigma_K (s)  \right]  , \qquad\;\; \sigma_i (s) = (1 - 4 M^2_i / s)^{1/2}\, .
\end{equation}
%
%
%

The different isospin components of the full amplitudes are given by $ T^I_{\pi \pi} (s) \equiv T^I_{D\to\pi \pi} (s) $ and $ T^I_{K K} (s) \equiv T^I_{D\to K K} (s) $.
At lowest order in weak interactions, the unitarity of the $S$ and $S_S$ matrices implies
%
\begin{equation}\label{eq:main_discontinuity_eq}
	\Sigma^{1/2} \, \begin{pmatrix}
	T^0_{\pi \pi} (s + i \epsilon) \\
	T^0_{K K} (s + i \epsilon) \\
	\end{pmatrix}\, =\,
	(S_S)^0_0 \;\, \Sigma^{1/2} \,
	\begin{pmatrix}
	T^0_{\pi \pi} (s - i \epsilon) \\
	T^0_{K K} (s - i \epsilon) \\
	\end{pmatrix}\, .
\end{equation}
We can decompose the full amplitudes as
\begin{equation}
	\begin{pmatrix}
	T^0_{\pi \pi} (s) \\
	T^0_{K K} (s) \\
	\end{pmatrix}\, =\,
	\Omega^{(0)} (s) \,
	\begin{pmatrix}
	T^{0\, (B)}_{\pi \pi} \\
	T^{0\, (B)}_{K K} \\
	\end{pmatrix}\, ,
\end{equation}
with the corresponding changes for $I=1,2$,
where $ T^{I\, (B)}_{\pi \pi} $ and $ T^{I\, (B)}_{K K} $ will be referred to as ``bare amplitudes'' (for which we will omit their possible $s$ dependence); they are polynomials in $s$ and may contain real zeros, while $ \Omega^{(I)} (s) $ has no zeros or poles.
As we will see, the bare amplitudes contain the CP-odd phases necessary to generate the CP asymmetries.
The rescattering part $ \Omega^{(I)} (s) $ of the transition amplitude satisfies then the following discontinuity equation:
\begin{equation}\label{eq:simple_DR}
    \Omega^{(I)} (s+i \epsilon)\, =\, \left[ {\bf 1} + 2 i\, T^I_0 (s)\, \Sigma (s) \right] \Omega^{(I)} (s-i \epsilon)\, \equiv\, \mathcal{S}^I (s)\; \Omega^{(I)} (s-i \epsilon)\, ,
\end{equation}
where $ \mathcal{S}^I (s) = {\bf 1} + 2 i T^I_0 (s)\, \Sigma (s) $, with $ \mathcal{S}^I (s) \mathcal{S}^I (s)^\ast = \mathcal{S}^I (s)^\ast \mathcal{S}^I (s) = {\bf 1} $.
This implies:
\begin{equation}
    {\rm Im} \, \Omega^{(I)} (s + i \epsilon)\, =\, T^{I \ast}_{0} (s)\, \Sigma (s)\, \Omega^{(I)} (s + i \epsilon)\, ,
\end{equation}
after using that $ \Omega^{(I)} (s-i \epsilon) ^\ast = \Omega^{(I)} (s+i \epsilon) $ (Schwarz reflection principle). In the following, we will drop ``$+i \epsilon$'' from $\Omega^{(I)} (s+i \epsilon)$.
The analyticity properties of $\Omega^{(I)} (s)$
guarantee that it satisfies the Cauchy integral relation:
\begin{equation}\label{eq:dispersive_equation}
	\Omega^{(I)} (s) = \frac{1}{\pi} \int^{\infty}_{4 M^2_\pi} \frac{T^{I \ast}_{0} (s') \Sigma (s') \Omega^{(I)} (s')}{s' - s - i \epsilon} d s' \, ;
\end{equation}
we will later adopt the normalization $\Omega^{(I)} (s_0) = {\bf 1}$, at a subtraction point $s_0$. In the two-channel coupled problem ($I=0$), we have:
%
\begin{equation}\label{eq:T00_expression}
	T^0_0 (s) = \begin{pmatrix}
	\frac{\eta^0_0 (s)\, e^{2 i \delta^0_0 (s)} - 1}{2 i\, \sigma_\pi (s)} && |g^0_0 (s)|\, e^{i \psi^0_0 (s)} \\
	|g^0_0 (s)|\, e^{i \psi^0_0 (s)} && \frac{\eta^0_0 (s)\, e^{2 i ( \psi^0_0 (s) - \delta^0_0 (s) )} - 1}{2 i\, \sigma_K (s)} \\
	\end{pmatrix}\, ,
\end{equation}
with the inelasticity parameter
\begin{equation}\label{eq:inelasticity_function_of_modg}
	\eta^0_0 (s) = \left[ 1 - 4\, \sigma_\pi (s)\, \sigma_K (s)\, | g^0_0 (s) |^2\, \Theta (s - 4 M^2_\pi)\, \Theta (s - 4 M^2_K) \right]^{1/2}\, .
\end{equation}
The sign of the off-diagonal elements of $ T^0_0 (s) $ is fixed at low energies by $\chi$PT \cite{Donoghue:1990xh}, given a choice of convention for the kaon pair isospin decomposition.



One can use that (Sokhotski-Plemelj relation):
\begin{equation}
	\frac{1}{x - x_0 - i \epsilon}\, =\, P \frac{1}{x - x_0} + i \pi\, \delta (x - x_0)
\end{equation}
to write alternatively:
\begin{equation}
	{\rm Re} [ \Omega^{(I)} (s) ] = \frac{1}{\pi} \dashint^{\infty}_{4 M^2_\pi} \frac{T^{I \ast}_{0} (s') \Sigma (s') \Omega^{(I)} (s')}{s' - s} d s' = \frac{1}{\pi} \left( \dashint^{4 M^2_K}_{4 M^2_\pi} + \; \dashint^{\infty}_{4 M^2_K} \right) \frac{ {\rm Im} [ \Omega^{(I)} (s') ] }{s' - s} d s'
\end{equation}
(the slashed integral represents its principal value).
Exploiting that the right-hand side is real, we get for the integration domain $ s' \geqslant 4 M_K^2 $ and any $ m \in \{ \pi, K \} $:
\begin{eqnarray}\label{eq:matrix_R_I}
	&& \begin{pmatrix}
	{\rm Re} [ (T^0_0)_{\pi \pi} ] \sigma_\pi & {\rm Re} [ (T^0_0)_{\pi K} ] \sigma_K \\
	{\rm Re} [ (T^0_0)_{K \pi} ] \sigma_\pi & {\rm Re} [ (T^0_0)_{K K} ] \sigma_K \\
	\end{pmatrix}
	\begin{pmatrix}
	{\rm Im} [ \Omega^{(0)}_{\pi m} ] \\ {\rm Im} [ \Omega^{(0)}_{K m} ]
	\end{pmatrix} \\
	&& \qquad\qquad = \begin{pmatrix}
	{\rm Im} [ (T^0_0)_{\pi \pi} ] \sigma_\pi & {\rm Im} [ (T^0_0)_{\pi K} ] \sigma_K \\
	{\rm Im} [ (T^0_0)_{K \pi} ] \sigma_\pi & {\rm Im} [ (T^0_0)_{K K} ] \sigma_K \\
	\end{pmatrix}
	\begin{pmatrix}
	{\rm Re} [ \Omega^{(0)}_{\pi m} ] \\ {\rm Re} [ \Omega^{(0)}_{K m} ]
	\end{pmatrix} \, .\nonumber
\end{eqnarray}
Admitting that the $ 2 \times 2 $ matrix on the left-hand side is invertible (the matrix $ T^0_0 \Sigma $ is invertible), then one can solve for $ {\rm Im} [ \Omega^{(0)} ] \equiv b $, which is plugged into the previous integral equation for the integration range $ s' \geqslant 4 M_K^2 $: indeed, this matrix equation can be written as $ \mathbf{R} \cdot b_m = \mathbf{I} \cdot a_m   \Leftrightarrow   b_m = \mathbf{R}^{-1} \cdot \mathbf{I} \cdot a_m $ if $ \mathbf{R} $ invertible, with obvious correspondence with Eq.~\eqref{eq:matrix_R_I}.
In the integration interval $ 4 M_\pi^2 \leqslant s' \leqslant 4 M_K^2 $ we have, like in the uncoupled case (and consider $ \psi^I_0 = \delta^I_0 \, \text{mod} \, \pi $ in this region):
\begin{equation}
	{\rm Re} [ (T^I_0)_{j \pi} ] \; {\rm Im} [ \Omega^{(I)}_{\pi m} ]\, =\, {\rm Im} [ (T^I_0)_{j \pi} ] \; {\rm Re} [ \Omega^{(I)}_{\pi m} ]\, .
\end{equation}
The adopted strategy is to solve for the real parts, and then use the previous relations to determine the imaginary parts.
Then:

\begin{equation}\label{eq:integral_equation}
	a_m = \frac{1}{\pi} \dashint^{4 M^2_K}_{4 M^2_\pi} d s' \frac{1}{s' - s} \, \begin{pmatrix}
	\tan \delta^0_0 (s') & 0 \\
	| g^0_0 (s') | \sigma_\pi (s') / \cos ( \delta^0_0 (s') ) & 0 \\ 
\end{pmatrix} \cdot a_m
+ \frac{1}{\pi} \dashint^{\infty}_{4 M^2_K} d s' \frac{\mathbf{R}^{-1} \cdot \mathbf{I} \cdot a_m}{s' - s}\, .
\end{equation}
%
%
We then solve for both $ a_\pi $ and $ a_K $,
and the final solution is:
\begin{equation}
	{\rm Re} [ \Omega^{(0)} ]\, =\, \left( a_\pi  \otimes  a_K \right) , \qquad\qquad {\rm Im} [ \Omega^{(0)} ]\, =\, \left( b_\pi  \otimes  b_K\right) \, ,
\end{equation}
where $ \otimes $ means that we combine the two dimension-2 vectors represented as columns into a $ 2 \times 2 $ matrix.
Note that $ \Omega^{(0)} (s_0) = \mathbf{1} $
implies that $ a_\pi $ and $ a_K $ are independent, and a similar comment applies for the imaginary parts $ b_\pi $ and $ b_K $.
The system of independent functions $ \chi^{(k)} (s) \equiv a_k + i \, b_k $ built from these real and imaginary parts is called a \textit{fundamental system of solutions} satisfying the discontinuity problem of Eq.~\eqref{eq:simple_DR}, see Ref.~\cite{Muskhelishvili}. There are $n$ such solutions in an $n$-channel coupled case.
A similar discussion holds for once-subtracted DRs. The use of subtracted DRs limits the dependence to high-energy data, which are typically less accurate or even missing; they may also be necessary in order to guarantee the convergence of the
dispersive integrals.


In the elastic limit, one can solve the integral equation explicitly \cite{Muskhelishvili,Omnes:1958hv}. Considering one subtraction:

\begin{equation} \label{eq:once_subtracted_Omnes_sol}
	\Omega^{(I)} (s) = \exp \left\{ i \, \delta^I_0 (s) \right\} \exp \left\{ \frac{s - s_0}{\pi} \, \dashint^\infty_{4 M^2_\pi} \frac{d z}{z - s_0} \frac{\delta^I_0 (z)}{z - s} \right\} \,,
\end{equation}
where $s_0 \le 4 M_\pi^2$ is the subtraction point, at which we have imposed $ \Omega^{(I)} (s_0) = 1 $.
The rightmost exponential above carries no zeros nor poles for well-behaved phase-shifts; it is obviously non-negative.
It is manifest that large phase-shifts are associated to large modulations of $ | \Omega^{(I)} (s) | $.
The phase-shift and the Omn\`{e}s factor $ | \Omega^{(I)} (s) | $ encode the effects of rescattering, and are necessary for a good qualitative and quantitative description of the transition amplitudes in the weak decay.
It is important stressing the universal character of this equation, which depends only on the phase-shift, and not the particular electroweak process under discussion. 


The previous equation leads to the following asymptotic behavior (see e.g. Ref.~\cite{Oller:2019rej}):

\begin{equation}\label{eq:index_behavior}
    \Omega^{(I)} (s) \to s^x \,, \qquad\qquad x = - \frac{\delta^I_J (\infty) - \delta^I_J (4 M_\pi^2)}{\pi}\, ,
\end{equation}
where at threshold $ \delta^I_J (4 M_\pi^2) = 0 $.
Therefore, if the Omn\`{e}s factor is supposed to vanish asymptotically, as it is expected when building form factors from the latter rescattering factor (see e.g. Ref.~\cite{Lepage:1980fj}), in the single-channel analysis one requires $ \delta^I_J (\infty) > \delta^I_J (4 M_\pi^2) $.


In the inelastic case, the determinant of $ \Omega^{(I)} (s) $ has an explicit analytical solution, from which a similar discussion holds. In the two-channel analysis for instance, Eq.~\eqref{eq:simple_DR} leads to:

\begin{equation}\label{eq:det_equation}
    \det \Omega^{(I)} (s+i \epsilon) = \exp \left\{ 2 \, i \, \psi^I_J (s) \right\} \det \Omega^{(I)} (s-i \epsilon)\, ,
\end{equation}
which does not depend on the inelasticity.\footnote{We note that we have not been able to find a function of the two-channel Omn\`{e}s matrix other than its determinant that does not depend on the inelasticity, for which there is an explicit analytical solution.}
A property of the fundamental system of solutions is that the individual indices $x_k$, describing the asymptotic behaviour of the fundamental solutions $\chi^{(k)} (s)$, do not depend on the particular choice of the fundamental system of solutions (see also next paragraph).
Their sum satisfies the relation \cite{Muskhelishvili}:
\begin{equation}\label{eq:sum_indices}
    \sum_{k = 1}^n x_k\, =\, x\, ,
\end{equation}
where $x$ is the index resulting from taking the determinant of Eq.~\eqref{eq:simple_DR} in the $n$-channel coupled analysis.
For instance, in the two-channel problem under discussion, $ x = - ( \psi^0_0 (\infty) - \psi^0_0 (4 M_\pi^2) )/\pi $.

Regarding the asymptotic behaviour, we comment on a specific case of later interest: if the sum of indices is $ x=-2 $, then one can have two independent solutions that vanish asymptotically and simultaneously, i.e., both having indices $ x_1 = -1 $ and $ x_2 = -1 $. If on the other hand the sum of indices is $ x=-3 $ for instance, then one can have two solutions that vanish asymptotically, i.e., $ x_i = -2 $ and $ -1 $, but they are not unique: to the solution that goes as $ -1 $ one can add a contribution from the one that goes as $ -2 $ (times a polynomial of degree up to one) and take this as the fundamental solution that replaces the previous one, while keeping the condition $\Omega^{(I)} (s_0)$ at a subtraction point $s_0$. In such cases, more physical information about the sought solutions has to be provided \cite{Babelon:1976kv}.


\subsection{Experimental inputs for the DRs}\label{sec:inputs_DRs}



Hereafter we discuss datasets and parameterizations of the inputs for DRs in isospin-zero and isospin-two. We point out the main qualitative features observed in phase-shifts and inelasticity, shown in Figs.~\ref{fig:phase_shifts}, \ref{fig:off_diagonal}, and \ref{fig:Inelasticity}.
We take the constrained fits to data (CFD) enforcing dispersive relations of Refs.~\cite{Kaminski:2006qe,GarciaMartin:2011cn,Pelaez:2019eqa,Pelaez:2020gnd}, that we discuss in more details below.


\subsubsection{Isospin-zero phase-shift of pion pairs}

\begin{figure}[t]
    \centering
    \includegraphics[scale=0.45]{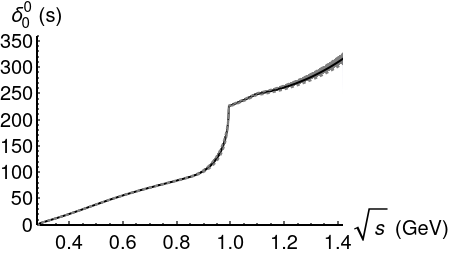} \hspace{3mm}
    \includegraphics[scale=0.45]{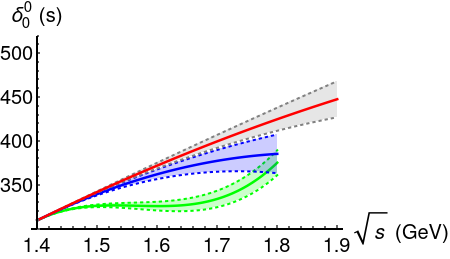} \\
    \vspace{3mm}
    \includegraphics[scale=0.45]{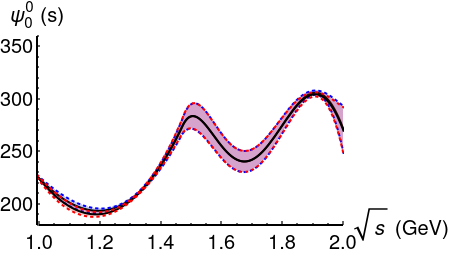} \hspace{3mm}
    \includegraphics[scale=0.45]{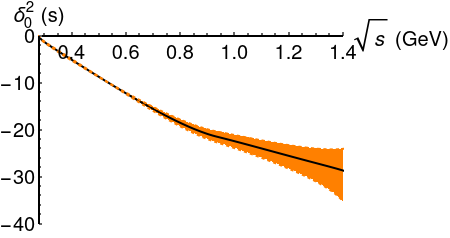}
    \caption{Set of phase-shifts from Refs.~\cite{GarciaMartin:2011cn,Pelaez:2019eqa,Pelaez:2020gnd,Kaminski:2006qe} used in our analysis. The $ \delta^0_0 (s) $ phase is shown in the
    upper panels for $2 M_\pi\leq \sqrt{s} \leq 1.42$~GeV (top, left) and $\sqrt{s} \geq 1.4$~GeV (top, right),
    where solutions~I (gray), II (blue) and III (green) are given. The $ \psi^0_0 (s) $ phase is shown in the (bottom, left), for solutions~B (blue) and C (red), which are very compatible; below the kaon pair threshold, $ \psi^0_0 (s) = \delta^0_0 (s) $. The $ \delta^2_0 (s) $ phase is shown in the (bottom, right), up to $\sqrt{s} = 1.4$~GeV and starting from the pion pair threshold. All phases are given in degrees.}
    \label{fig:phase_shifts}
\end{figure}

We use the analyses of Refs.~\cite{GarciaMartin:2011cn,Pelaez:2019eqa}.
As seen from the top-left panel of Fig.~\ref{fig:phase_shifts},
the phase-shift starts at zero at the pion pair production threshold,
and shows a steady increase sufficiently below the threshold for kaon pair production, due to the presence of the $\sigma$ resonance, which is located deep into the first Riemann sheet, away from the real axis.
Then, there is a quick increase of the phase-shift, due to the presence of the $f_0 (980)$ resonance, which is relatively narrow. Another analytical feature in the region $\sim 1$~GeV is the threshold for kaon pair production.
Subsequently, the phase-shift grows steadily; in this energy region there exist the well-established resonances $f_0 (1370)$, $f_0 (1500)$ and $f_0 (1710)$, which to some extent overlap among themselves (for a recent discussion of $f_0 (1370)$, see Ref.~\cite{Pelaez:2022qby}).

Above around $1.42$~GeV, Ref.~\cite{Pelaez:2019eqa} considers different datasets, which are not consistent among themselves, providing purely descriptive phase-shift parameterizations separately for each of them, see the top-right panel of Fig.~\ref{fig:phase_shifts}. Solution~I \cite{Kaminski:1996da,CERN-Cracow-Munich:1978kaq,Hyams:1973zf,Grayer:1974cr} follows from a dataset that extends up to $E_0 = 1.9$~GeV, while the datasets leading to solutions~II \cite{Hyams:1975mc} and III \cite{Ochs:2013gi} extend up to $E_0 = 1.8$~GeV.

\subsubsection{Isospin-zero phase-shift of kaon pairs}

We consider the combined analysis of $\pi\pi \to KK$ and $K\pi \to K\pi$ employing crossing symmetry of Ref.~\cite{Pelaez:2020gnd}, see also Ref.~\cite{Pelaez:2018qny}. There are two possible solutions, B and C, that are well compatible, see the bottom-left panel of Fig.~\ref{fig:phase_shifts} (see also comments below).
The curve extends up to $E_0 = 2$~GeV.
There is a clear structure in the phase-shift in the region $1.2 - 2$~GeV, that might be in part due to the isoscalar-scalar resonances mentioned above, with the phase-shift decreasing at times.\footnote{There is an interesting result in Quantum Mechanics, according to which the phase-shift cannot decrease too quickly in order to respect causality, see \cite{Wigner:1955zz} and e.g. Ref.~\cite{Newton,weinberg_2015}. In the present situation, we observe that $ - 2 \, \hbar \, c \, \frac{d \psi^0_0 (E^2)}{d E} \lesssim 4 $~fm, which gives a crude estimate of the minimum range of the potential as required from the requirement of causality.}

\subsubsection{Isospin-zero inelasticity}

\begin{figure}[t]
    \centering
    \includegraphics[scale=0.35]{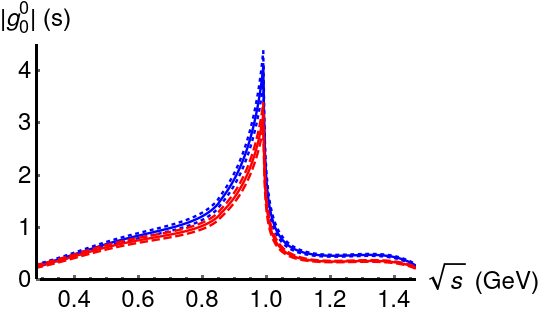} \hspace{3mm}
    \includegraphics[scale=0.42]{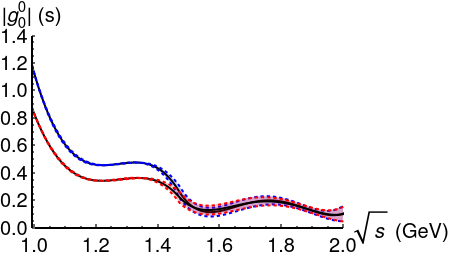}
    \caption{Off-diagonal $T$-matrix element from Ref.~\cite{Pelaez:2020gnd} for solutions~B (blue) and C (red). The two sets, valid along the energy ranges from the pion pair threshold and up to $1.47$~GeV (left) and from the the kaon pair threshold and up to $2$~GeV (right), are combined according to the procedure described in the text.}
    \label{fig:off_diagonal}
\end{figure}

\begin{figure}[t]
    \centering
    \includegraphics[scale=0.25]{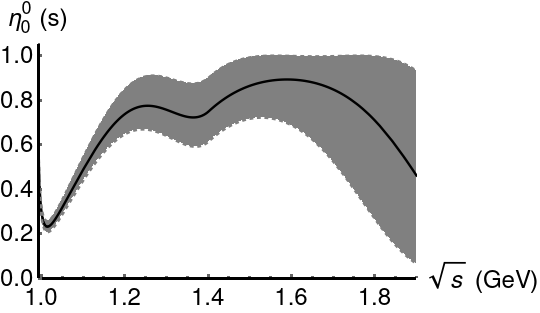} \hspace{1mm}
    \includegraphics[scale=0.25]{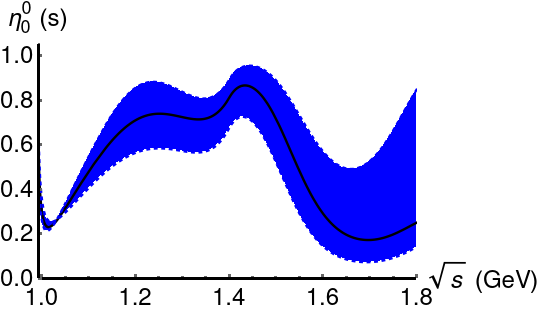} \hspace{1mm}
    \includegraphics[scale=0.25]{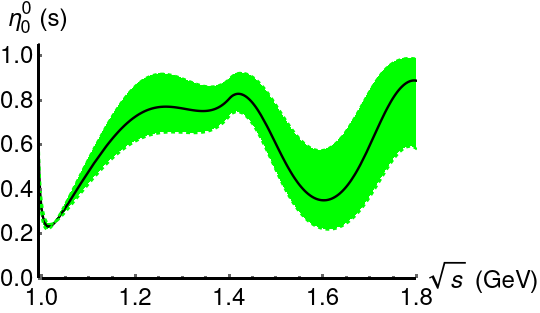}
    \caption{Inelasticity from Ref.~\cite{Pelaez:2019eqa} extracted from pion pair rescattering. Below the kaon pair threshold, $ \eta^0_0 (s) = 1 $. Three solutions are shown, namely, solutions~I (left), II (center) and III (right).}
    \label{fig:Inelasticity}
\end{figure}

Below the threshold for their on-shell production, virtual kaon pairs produce off-diagonal elements in the two-channel rescattering matrix, with their impact seen in the first term in the right-hand side of Eq.~\eqref{eq:integral_equation}. They do not produce an absorptive part though, i.e., do not alter the evolution of the phase motion, and the off-diagonal phase-shift therein is then the one observed in pion pair rescattering.
Note, however, that it does not mean that the inelasticity below the kaon pair threshold varies, being $\eta^0_0 = 1$ below this threshold.
We consider Refs.~\cite{Pelaez:2020gnd,ArkaitzRodas} for a parameterization of such effect, illustrated in the left panel of Fig.~\ref{fig:off_diagonal} (for definiteness, when not specified we employ sol.~B), see also Ref.~\cite{Pelaez:2018qny}.

The inelasticity $ \eta^0_0 $ can be extracted from the off-diagonal $T$-matrix element $ | g^0_0 | $ via a combined analysis of $\pi\pi \to KK$ and $K\pi \to K\pi$, and is available up to $E_0 = 2$~GeV, as illustrated in the right panel of Fig.~\ref{fig:off_diagonal}. This leads to two solutions, B \cite{Longacre:1986fh} and C \cite{Cohen:1980cq,Etkin:1981sg}, corresponding to inconsistent datasets below $\sim 1.47$~GeV, and thus their parameterizations of the inelasticity differ substantially below that point.
We combine the effect generated by off-shell kaon pairs \cite{Pelaez:2020gnd,ArkaitzRodas} with the explicit parameterizations found in Ref.~\cite{Pelaez:2020gnd} valid above $2 M_K$. The two sets of curves are combined at a matching point of $\sqrt{1.2}$~GeV \cite{ArkaitzRodas}, and the corresponding solutions will be called B' and C' in the following. There is a very small discontinuity at the matching scale (of $9$\% for solution~B', and of $8$\% for solution~C').
Right above the kaon pair threshold and below the matching scale, there is a short window in which the unitarity bound is violated, manifested as the impossibility of defining a real inelasticity therein via the use of Eq.~\eqref{eq:inelasticity_function_of_modg}. However, this corresponds to a tiny region (long of $\sim 30$~MeV for solution~B', and of $\sim 10$~MeV for solution~C'), in which we set the inelasticity to zero.

An alternative for the extraction of the inelasticity $ \eta^0_0 $ is to look directly at the rescattering process $\pi\pi \to \pi\pi$.
The extraction of the parameterization for the phase-shift of the pion pair system discussed above is done simultaneously to the extraction of the parameterization for the inelasticity, for which then we also have three solutions \cite{Pelaez:2019eqa}, illustrated in Fig.~\ref{fig:Inelasticity}. As before, solution~I extends up to $E_0 = 1.9$~GeV, while solutions~II and III extend up to $E_0 = 1.8$~GeV.
(We correct typos found in Ref.~\cite{Pelaez:2019eqa}, namely: $ \epsilon_4 $ in their Tab.~2 comes with the wrong sign, and $K_i$ must be multiplied by $ M_K^2 $ \cite{ArkaitzRodas}.)
Solution~III shows a sharp dip in the inelasticity around $\sim 1.6$~GeV, and a distinguishing phase motion compared to the other two solutions, which may signal the presence of the resonance $f_0 (1500)$.
As further discussed later, inelasticities extracted in this way carry large uncertainties.
At the energy $2 M_K$, the off-diagonal element resulting from this inelasticity is combined with the one from Refs.~\cite{Pelaez:2020gnd,ArkaitzRodas}, that describes off-shell kaon pairs. This produces an abrupt change across a few MeV of the off-diagonal $T$-matrix element at about $2 M_K$, which is expected to have a limited impact on the fundamental Omn\`{e}s solutions far away from this value of the energy.
Moreover, combining the two curves at $2 M_K$ generates a consistent trend, since below (above) $2 M_K$ the modulus of the off-diagonal $T$-matrix is decreasing (increasing) quickly with increasing (decreasing) energies.

The different sets of inelasticities discussed above, solutions~I-III and solutions~B' and C', display important differences. In discussing solutions of the dispersive equations, we will show results for each of them separately.

\subsubsection{Isospin-two phase-shift and inelasticity of pion pairs}\label{sec:isospin_two}


The phase-shift starts at zero at the pion pair production threshold, and decreases steadily in the region extending up to about $1.45$~GeV \cite{Cohen:1973yx,Losty:1973et,Hoogland:1977kt}. A parameterization is provided by Refs.~\cite{Kaminski:2006qe,JoseRPelaez}, see the bottom-right panel of Fig.~\ref{fig:phase_shifts}, which does not include Ref.~\cite{Durusoy:1973aj}.
The extracted inelasticity is close to the elastic limit in that energy range \cite{Kaminski:2006qe}.

Although to our knowledge a parameterization is not available (in particular, taking into account dispersive constraints), Ref.~\cite{Durusoy:1973aj} extracts data up to $2.2$~GeV. It is seen that the phase-shift has the tendency to decrease up to $1.3$~GeV, and then to increase subsequently. At around $M_D$, the phase-shift equals a few negative degrees, but carries a large uncertainty. There seems to be a large inelasticity at around $M_D$, although again large uncertainties are present. (This overall behaviour of the phase-shift can be reproduced via an elastic analysis relying on $\chi$PT and Resonance Chiral Theory (R$\chi$T), with $t$-channel exchange of $\rho$, etc.; see also Ref.~\cite{Zou:2004qn}.)

As discussed above around Eq.~\eqref{eq:index_behavior}, the phase-shift in an elastic analysis should become positive (vanish) so that the Omn\`{e}s solution goes to zero (respectively, a constant) at infinity. 
This requires some underlying physics to change the sign of the isospin-two phase-shift, such as the presence of a resonance.
We also note that no distinct feature is seen in the isospin-two $\pi\pi\to\pi\pi$ study of Ref.~\cite{Caprini:2011ky}, for which however contributions to the cross-section other than the $S$-wave become increasingly important at higher energies.

We will later in the text extract the Omn\`{e}s factor $| \Omega^{(2)} |$ from the branching ratio of the charged decay mode $D^+ \to \pi^0 \pi^+$, and vary the isospin-two phase-shift to reproduce the $D^0 \to \pi^- \pi^+, \pi^0 \pi^0$ branching ratios.
We reserve further discussion about the isospin-two inelasticity to future work \cite{PSVS_Ieq2:2023}. 





\subsection{Solutions of the coupled channel DRs}\label{sec:sols_coupled_channel_DRs}


To employ the DRs,
we extrapolate the phase-shift and inelasticity curves discussed above beyond their indicated endpoints $E_0 = 1.8-2$~GeV \cite{Moussallam:1999aq}:
\begin{equation}\label{eq:extending_phase_shifts}
	\delta_0^0 (E) = n^\ast \pi + (\delta_0^0 (E_0) - n^\ast \pi) f_\delta \left( \frac{E}{E_0} \right) , \quad \delta_K (E) = \ell^\ast \pi + (\delta_K (E_0) - \ell^\ast \pi) f_\delta \left( \frac{E}{E_0} \right) ,
\end{equation}
where the chosen $ f_\delta (x) = 2 / (1 + x^{m^\ast_\delta}) $ has the virtue of being a smooth function connecting the values at the endpoints to the asymptotic values (Ref.~\cite{Moussallam:1999aq} takes $m^\ast_\delta = 3$; 
we note that the asymptotic behaviour of the phase-shift involved in the vector form factor of the pion is discussed in Ref.~\cite{Leutwyler:2002hm}). The values of $n^\ast + \ell^\ast \geq 2$ ensure that at least one of the fundamental solutions tends to zero at infinite energies. We take $n^\ast, \ell^\ast$ as integer values (as it results from having resonant effects; i.e., we neglect non-resonant effects for this sake).
Then, we set $\ell^\ast = -1$ since $ \delta_K (E) \equiv \psi^0_0 (E) - \delta^0_0 (E) $ is close to $- \pi$ at $E_0$.
Finally, we take $n^\ast = 3$.
Moreover, it suffices to ensure the good behaviour of the fundamental solutions, as it leads in practice to two independent solutions of indices $x_{1,2} = -1$.
These solutions are uniquely determined, after specifying the condition $ \Omega^{(0)} (s_0) = {\bf 1} $ at the subtraction point $s_0$.



A similar extrapolation is taken for the inelasticity:

\begin{equation}
    \eta^0_0 (E) = \eta_\infty + (\eta^0_0 (E_0) - \eta_\infty) f_\eta \left( \frac{E}{E_0} \right) \,.
\end{equation}
Its limiting value is set to $\eta_\infty = 1$.
Together with the limit values of the phase-shifts, these conditions satisfy the asymptotic behaviour discussed in Ref.~\cite{Warnock1967}.
Large values of $m_\eta^\ast$ (i.e., $ \eta^0_0 $ approaching faster its asymptotic value) would require some underlying dynamics, such as the appearance of resonances not yet firmly established \cite{Workman:2022ynf}, and for this reason we later display only values in the range $ m_\eta^\ast \in \{ 1, 2, 3 \} $.

To full generality, there is no known explicit solution in the inelastic case.
The numerical method used is described in App.~\ref{app:num_sol_DRs} (we discuss an explicit solution valid under certain conditions in App.~\ref{app:analytic_solution}), and relies on the parameterization of data previously discussed.\footnote{In the $N/D$ method, phase-shift parameterizations and Omn\`{e}s functions are extracted simultaneously in the fits to the rescattering data \cite{Danilkin:2020pak,Deineka:2022izb}.}

A sample of typical 
Omn\`{e}s matrices is provided in Tab.~\ref{tab:num_sols_Omnes} for various scenarios:
columns correspond to solutions I-III for the phase-shifts, and also inelasticity;
the first block of rows corresponds to the inelasticity directly extracted from $ \pi \pi \to \pi \pi $, while the second block of rows corresponds to the inelasticity calculated from the off-diagonal $T$-matrix element as in Eq.~\eqref{eq:inelasticity_function_of_modg}, for which there are solutions B' and C'.
We observe a strong dependence of the Omn\`{e}s solution with the inelasticity employed,
which in the case of the first block of rows carries a large uncertainty. Varying the latter uncertainties leads to profiles $ \eta^0_0 - \delta \eta^0_0 $, which seek to saturate the error bars attached to the inelasticities found in Ref.~\cite{Pelaez:2019eqa} towards smaller values of $ \eta^0_0 $.\footnote{Possible correlations among the different uncertainties for phase-shifts and inelasticity are neglected here.}
In a companion paper, we provide a discussion of CP asymmetries that does not depend on the input employed for the inelasticity \cite{PSVS:2023}.



When calculating the Omn\`{e}s matrices, we verify that $ \Omega^{(0)}_{11} (M_K^2) $ is in good agreement with a similar calculation relying on an elastic analysis \cite{Pallante:1999qf,Pallante:2000hk}: were there a sizable difference, it would spoil the good comparison with the $\chi$PT calculation of this same quantity.

\begin{sidewaystable}
    \centering
    \renewcommand{\arraystretch}{1.8}
    \begin{tabular}{|c||c|c|c|}
    \hline
         & solution~I & solution~II & solution~III \\
    \hline\hline 
        $ \eta^0_0 $, $ m^\ast_\eta = 1 $ & $ \Omega^{(0)} = \begin{pmatrix}
        0.80 \, e^{+1.60 \, i} & 1.01 \, e^{-1.69 \, i} \\
        0.56 \, e^{-1.50 \, i} & 0.59 \, e^{+2.07 \, i} \\
	\end{pmatrix} $
        & $ \Omega^{(0)} = \begin{pmatrix}
        0.39 \, e^{+1.64 \, i} & 0.59 \, e^{-2.13 \, i} \\
        0.51 \, e^{-1.31 \, i} & 0.56 \, e^{+2.43 \, i} \\
	\end{pmatrix} $
        & $ \Omega^{(0)} = \begin{pmatrix}
        0.71 \, e^{+0.53 \, i} & 1.35 \, e^{-2.67 \, i} \\
        0.38 \, e^{-0.98 \, i} & 0.42 \, e^{+2.65 \, i} \\
	\end{pmatrix} $ \\
    \hline 
        $ \eta^0_0 - \delta \eta^0_0 $, $ m^\ast_\eta = 1 $ & $ \Omega^{(0)} = \begin{pmatrix}
        0.56 \, e^{+1.84 \, i} & 0.61 \, e^{-1.73 \, i} \\
        0.57 \, e^{-1.41 \, i} & 0.58 \, e^{+2.25 \, i} \\
	\end{pmatrix} $
        & $ \Omega^{(0)} = \begin{pmatrix}
        0.42 \, e^{+1.75 \, i} & 0.54 \, e^{-2.05 \, i} \\
        0.51 \, e^{-1.33 \, i} & 0.55 \, e^{+2.43 \, i} \\
	\end{pmatrix} $
        & $ \Omega^{(0)} = \begin{pmatrix}
        0.35 \, e^{+1.13 \, i} & 0.74 \, e^{-2.47 \, i} \\
        0.50 \, e^{-1.18 \, i} & 0.55 \, e^{+2.48 \, i} \\
	\end{pmatrix} $ \\
    \hline 
        $ \eta^0_0 - \delta \eta^0_0 $, $ m^\ast_\eta = 2 $ & $ \Omega^{(0)} = \begin{pmatrix}
        0.58 \, e^{+1.80 \, i} & 0.64 \, e^{-1.74 \, i} \\
        0.58 \, e^{-1.37 \, i} & 0.61 \, e^{+2.26 \, i} \\
	\end{pmatrix} $
        & $ \Omega^{(0)} = \begin{pmatrix}
        0.43 \, e^{+1.64 \, i} & 0.58 \, e^{-2.10 \, i} \\
        0.52 \, e^{-1.25 \, i} & 0.57 \, e^{+2.48 \, i} \\
	\end{pmatrix} $
        & $ \Omega^{(0)} = \begin{pmatrix}
        0.40 \, e^{+1.01 \, i} & 0.80 \, e^{-2.50 \, i} \\
        0.50 \, e^{-1.11 \, i} & 0.56 \, e^{+2.53 \, i} \\
	\end{pmatrix} $ \\
     \hline 
        $ \eta^0_0 - \delta \eta^0_0 $, $ m^\ast_\eta = 3 $ & $ \Omega^{(0)} = \begin{pmatrix}
        0.60 \, e^{+1.76 \, i} & 0.66 \, e^{-1.74 \, i} \\
        0.60 \, e^{-1.33 \, i} & 0.63 \, e^{+2.26 \, i} \\
	\end{pmatrix} $
        & $ \Omega^{(0)} = \begin{pmatrix}
        0.44 \, e^{+1.53 \, i} & 0.61 \, e^{-2.16 \, i} \\
        0.52 \, e^{-1.17 \, i} & 0.59 \, e^{+2.53 \, i} \\
	\end{pmatrix} $
        & $ \Omega^{(0)} = \begin{pmatrix}
        0.45 \, e^{+0.91 \, i} & 0.86 \, e^{-2.53 \, i} \\
        0.50 \, e^{-1.04 \, i} & 0.57 \, e^{+2.58 \, i} \\
	\end{pmatrix} $ \\
    \hline\hline 
        sol.~B': $ | g^0_0 | $ & $ \Omega^{(0)} = \begin{pmatrix}
        2.01 \, e^{+1.39 \, i} & 2.47 \, e^{-1.76 \, i} \\
        0.37 \, e^{-0.33 \, i} & 0.54 \, e^{+3.05 \, i} \\
	\end{pmatrix} $
        & $ \Omega^{(0)} = \begin{pmatrix}
        1.91 \, e^{+0.60 \, i} & 2.78 \, e^{-2.55 \, i} \\
        0.31 \, e^{-0.23 \, i} & 0.45 \, e^{+3.30 \, i} \\
	\end{pmatrix} $
        & $ \Omega^{(0)} = \begin{pmatrix}
        2.20 \, e^{+0.43 \, i} & 3.55 \, e^{-2.72 \, i} \\
        0.35 \, e^{+0.03 \, i} & 0.57 \, e^{+3.40 \, i} \\
	\end{pmatrix} $ \\
    \hline 
        sol.~C': $ | g^0_0 | $ & $ \Omega^{(0)} = \begin{pmatrix}
        1.83 \, e^{+1.38 \, i} & 2.65 \, e^{-1.76 \, i} \\
        0.34 \, e^{-0.40 \, i} & 0.57 \, e^{+3.00 \, i} \\
	\end{pmatrix} $
        & $ \Omega^{(0)} = \begin{pmatrix}
        1.80 \, e^{+0.59 \, i} & 3.11 \, e^{-2.56 \, i} \\
        0.29 \, e^{-0.24 \, i} & 0.49 \, e^{+3.24 \, i} \\
	\end{pmatrix} $
        & $ \Omega^{(0)} = \begin{pmatrix}
        2.09 \, e^{+0.43 \, i} & 3.94 \, e^{-2.72 \, i} \\
        0.32 \, e^{-0.03 \, i} & 0.61 \, e^{+3.34 \, i} \\
	\end{pmatrix} $ \\
    \hline
    \end{tabular}
    \caption{Sample of Omn\`{e}s solutions at $ s = M_D^2 $. In the main text, the case of solution~I for the phase-shift and the inelasticity $ \eta^0_0 - \delta \eta^0_0 $, with $ m^\ast_\eta = 2 $, is referred to as the reference case.}
    \label{tab:num_sols_Omnes}
\end{sidewaystable}

\subsection{Partial-wave transition amplitudes}\label{sec:subtraction_constants}




In order to build transition amplitudes from the rescattering effects encoded in $ \Omega^{(0)} (s) $ (or analogously for isospin-one and -two, that we treat elastically), we need to specify the polynomial ambiguities in $T_{\pi \pi}^{0\, (B)}$ and $ T_{K K}^{0\, (B)}$ of the once-subtracted DRs.
Summing over the possible solutions to the two-channel coupled analysis problem, times subtraction constants, we have:

\begin{equation}
	\begin{pmatrix}
	T_{\pi \pi}^0 (s) \\
	T_{K K}^0 (s) \\
	\end{pmatrix} \, =\,
	\begin{pmatrix}
	\Omega^{(0)}_{11} (s) & \Omega^{(0)}_{12} (s) \\
	\Omega^{(0)}_{21} (s) & \Omega^{(0)}_{22} (s) \\
	\end{pmatrix}
	\begin{pmatrix}
	T_{\pi \pi}^{0\, (B)} \\
	T_{K K}^{0\, (B)} \\
	\end{pmatrix}\, ,
\end{equation}
where since we deal with charmed-meson decays, $s \to M_D^2$.

The polynomials $T_{\pi \pi}^{0\, (B)}$ and $T_{K K}^{0\, (B)}$ are fixed by physical considerations relying on a large-$ N_C $ expansion.
In the limit $N_C\to\infty$, the scattering phase-shifts are exactly zero and, therefore, $\Omega^{(I)}(s) = {\bf 1}$. Moreover, in this limit the hadronic matrix elements of the short-distance four-quark operators factorize into matrix elements of QCD currents. The bare amplitudes $T_{\pi \pi}^{0\, (B)}$ and $T_{K K}^{0\, (B)}$ correspond then
to tree insertions of different local operators, current-current and 
penguin ones, while
topologies beyond trees are generated via rescattering effects.
The factorized expressions are written in terms of decay constants 
and form factors (e.g., $ D \to \pi $, or $ D \to K $), given in Apps.~\ref{app:norm_sign_convs} and \ref{app:matrix_elements}.
It follows from the present discussion that the subtraction constants require perturbative and non-perturbative elements: decay constants,
form factors, and Wilson coefficients.
As it has been discussed,
rescattering is taken into account dispersively, and it is in fact
suppressed in the large-$N_C$ counting.
Decay constants and form factors also integrate non-perturbative QCD effects that, although sub-leading in the large-$N_C$ counting, are not included in the rescattering matrix $\Omega^{(I)}(s)$.
Note that the resulting subtraction constants are real (in the CP-conserving limit), strong complex phases being developed in the rescattering.
In the context of $ K \to \pi \pi $ transitions, the polynomial ambiguities can be determined via $\chi$PT \cite{Pallante:1999qf,Pallante:2000hk}.
(For a discussion of form factors built from the same rescattering effects, their asymptotic behaviour, and the use of $\chi$PT to determine the subtraction constants, see e.g. Refs.~\cite{Donoghue:1990xh,Guerrero:1997ku,Moussallam:1999aq,Pich:2001pj,Celis:2013xja}.)


The subtraction point is taken at $s_0 = M_\pi^2$, as suggested by $T_{\pi \pi}^{0\, (B)} \propto M_D^2 - M_\pi^2$. At this point, $\Omega^{(0)}$ is set to the identity ${\bf 1}$. Any modulation of $\Omega^{(0)}$ above $s_0$ results then from rescattering effects.
We observe, however, a very small dependence with the choice of the subtraction point, as seen from the two following solutions:

\begin{eqnarray}
&& \Omega^{(0)} (M_D^2) = \left(
\begin{array}{cc}
 0.59 \, e^{+1.81 \, i} & 0.64 \, e^{-1.74 \, i} \\
 0.59 \, e^{-1.38 \, i} & 0.62 \, e^{+2.26 \, i} \\
\end{array}
\right) \,, \qquad s_0 = 0\, , \\
%
%
%
%
&& \Omega^{(0)} (M_D^2) = \left(
\begin{array}{cc}
 0.57 \, e^{+1.71 \, i} & 0.61 \, e^{-1.72 \, i} \\
 0.56 \, e^{-1.27 \, i} & 0.58 \, e^{+2.24 \, i} \\
\end{array}
\right) \,, \qquad s_0 = 4 \, M_\pi^2\, ,
\end{eqnarray}

\noindent
which are calculated with the same inputs as used for the so-called reference case of Tab.~\ref{tab:num_sols_Omnes} to be discussed below, but with different subtraction points.
Moreover, given that $ M_\pi^2, M_K^2 \ll M_D^2 $, we observe a very small numerical impact of keeping the masses of the light mesons with respect to neglecting them in the expressions of the physical observables.

\section{Theoretical predictions}\label{sec:physical_predictions}



%
%


Before moving to the numerical predictions for branching ratios and CP asymmetries, we first discuss the mechanisms at play responsible for generating a non-vanishing level of CP violation. Detailed technical discussions are given in a series of appendices:
App.~\ref{app:norm_sign_convs} discusses the relevant decay constants and form factors, App.~\ref{app:matrix_elements} gives the expressions for the bare decay amplitudes, and
the isospin decomposition of the transition matrix elements is detailed in App.~\ref{app:isospin_decomposition}.


\subsection{Mechanisms of CP violation}\label{sec:mechanism_CPV}



We consider tree insertions of the short-distance operator basis provided in Eq.~\eqref{eq:operator_list}, whose matrix elements can be found in App.~\ref{app:matrix_elements}.
The CP-violating effects are generated through the interference of amplitudes with different weak and strong phases. Let us consider first the isospin-zero decay amplitudes that exhibit the CKM structure displayed in Eq.~(\ref{eq:H_eff}):
\begin{equation}
\label{eq:CP-structure}
 \begin{pmatrix} T^0_{\pi\pi} (s) \\ T^0_{KK} (s) \end{pmatrix}
 \, =\,  \Omega^{(0)} (s) \; 
 \begin{pmatrix} \lambda_d \, T^{CC}_{\pi \pi} - \lambda_b \, T^{P}_{\pi \pi}
  \\ \lambda_s \, T^{CC}_{K K} - \lambda_b \, T^{P}_{K K} \end{pmatrix}
  \,\equiv\,
\begin{pmatrix} \mathcal{A}_0^{\pi} + i\, \mathcal{B}_0^{\pi}
\\ \mathcal{A}_0^{K} + i\, \mathcal{B}_0^{K} \end{pmatrix}\, .  
\end{equation}
The uncoupled $I=1$ and $2$ amplitudes can also be written in a similar (simpler) way. The CP-even strong phases are generated by the rescattering matrices $\Omega^{(I)}$, while the CP-odd weak phases originate in the CKM factors $\lambda_q$ appearing in the bare amplitudes, which are different for charged-current ($T^{CC}_{\pi\pi,KK}$) and penguin ($T^{P}_{\pi\pi,KK}$) operators. We have decomposed the full decay amplitudes into their CP-even ($\mathcal{A}_I$) and CP-odd ($\mathcal{B}_I$) components. Obviously, the $\mathcal{A}_I$ amplitudes depend on the parameters $\text{Re} \{ \lambda_q \}$, while $\mathcal{B}_I$
are governed by $\text{Im} \{ \lambda_q \}$. 
Despite the different sizes of their corresponding Wilson coefficients, $ T^{CC}_{\pi\pi,KK} \sim T^{P}_{\pi\pi,KK}$ due to the large pre-factors coming with $Q_6$ insertions, see App.~\ref{app:matrix_elements}.


%

Observable effects must be stated in terms of rephasing-invariant quantities. Other than the quartets
\begin{equation}
    Q_{\alpha i \beta j} \equiv V_{\alpha i} V_{\beta j} V_{\alpha j}^\ast V_{\beta i}^\ast \,,
\end{equation}
rephasing-invariant objects also include the moduli of the elements of the CKM matrix; for a review, see Ref.~\cite{Branco:1999fs}. 
The 
relevant rephasing-invariant quantities have the following numerical values:
\begin{eqnarray}
	Q_{u d c b} &\!\! =&\!\! - A^2 \lambda^6 (\rho + i \eta) + \mathcal{O} (\lambda^8) \,\simeq\, - (1.3 + i \, 3.1) \times 10^{-5} \,, \\
	Q_{u d c s}&\!\! =&\!\! - \lambda^2 + \lambda^4 + A^2 \lambda^6 (1 - \rho + i \eta) + \mathcal{O} (\lambda^8) \,\simeq\, -0.048 + i \, 3.1 \times 10^{-5} \,, \\
	Q_{u s c b}&\!\! =&\!\! A^2 \lambda^6 (\rho + i \eta) + \mathcal{O} (\lambda^8) \,\simeq\, (1.3 + i \, 3.1) \times 10^{-5}\, ,
\end{eqnarray}
and
\begin{eqnarray}
	 | \lambda_d |^2 &\!\! =&\!\! \lambda^2 + \mathcal{O} (\lambda^4) \,\simeq\, 0.051 \,, \\
	 | \lambda_s |^2 &\!\! =&\!\! \lambda^2 + \mathcal{O} (\lambda^4) \,\simeq\, 0.051 \,, \\
	 | \lambda_b |^2 &\!\! =&\!\! A^4 \lambda^{10} (\rho^2 + \eta^2) + \mathcal{O} (\lambda^{12}) \,\simeq\, 2.3 \times 10^{-8}\, .
\end{eqnarray}
Note that $ \lambda_d $ and $ \lambda_s $ cannot be chosen simultaneously real, since the quartets $ Q_{\alpha i \beta j} $ are rephasing-invariant and $ Q_{u d c s} = V_{u d} V_{c s} V^\ast_{u s} V^\ast_{c d} = \lambda_d \lambda_s^\ast $. This is particularly important in the presence of rescattering effects, under which the isoscalar amplitudes depend on both, $ \lambda_d $ and $ \lambda_s $. (Numerical values of $ \text{Re} \{ \lambda_q \} $ and $ \text{Im} \{ \lambda_q \} $, $ q = d, s, b $, in the usual convention for the CKM matrix elements are found in App.~\ref{app:num_inputs}.)


Thus, the rescattering of the final pseudoscalar mesons generates a pure $I=0$ contribution to the CP asymmetries, originating in the interference of the intermediate $\pi\pi$ and $KK$ contributions.
Written in a rephasing-invariant way, the full contribution of isospin-zero--only amplitudes to the numerator of the direct CP asymmetries is given by 
\begin{equation}\label{eq:CPasym_numerator}
\mathrm{num} \, (A_{CP}^i)_{I=0}
\, = \,
4 \, \omega^{({\rm Im})}_i \, J \, \left( T^{C C}_{\pi \pi} \, T^{C C}_{K K}
	+ T^{C C}_{\pi \pi} \, T^{P}_{K K}
	+ T^{P}_{\pi \pi} \, T^{C C}_{K K} \right) \,.
\end{equation}

\noindent
This contribution is governed by the Jarlskog parameter
$ J = \text{Im} \{ Q_{u d c s} \} = r_{\rm CKM}\, | \lambda_d |^2 $, where $ r_{\rm CKM} \equiv \text{Im} \{ \lambda_b/\lambda_d \} $, and the dynamical rescattering factors
\begin{equation}
 \omega^{({\rm Im})}_i \,\equiv\, \text{Im} \{ \Omega^{(0) \ast}_{i 1} \Omega^{(0)}_{i 2} \} \, .   
\end{equation}
%
%
%
The quantity $ \omega^{({\rm Im})}_\pi \equiv \omega^{({\rm Im})}_1 $ ($ \omega^{({\rm Im})}_K \equiv \omega^{({\rm Im})}_2 $) controls the amount of CP violation in $ D^0 \to \pi \pi $ (respectively, $ D^0 \to K K $) coming exclusively from the interference of isospin-zero contributions.
The possibility of having a source of CP violation coming exclusively from isospin-zero amplitudes has been pointed out by, e.g., Ref.~\cite{Franco:2012ck}.
Such a case is not possible in kaon decays, since the dynamics therein is elastic.

The source of CP violation coming from current-current operators, due to the non-unitarity of the $ 2 \times 2 $ CKM sub-matrix, and the suppression of contributions from penguin operators due to small Wilson coefficients, have often been pointed out in the literature, see e.g. Refs.~\cite{Brod:2012ud,Khodjamirian:2017zdu}.
Note, however, that in Ref.~\cite{Soni:2019xko} the quantity analogous to $T^{P}_{\pi\pi,KK}$ 
generates the needed CP-odd amplitude, in a mechanism in which the operators $Q_{5,6}$ couple $D^0$ to $f_0 (1710)$, that subsequently decays to pion and kaon pairs. The state $f_0 (1710)$ being close to be on-shell, it can produce some enhancement of the amplitudes, and (part of) the strong phases come from the absorptive part of the $f_0 (1710)$ lineshape; see also Ref.~\cite{Schacht:2021jaz}.
We note that the imprints of resonances should manifest in the phase-shifts and inelasticity that are the inputs of the DRs discussed previously.

%

The full contribution of isospin-zero--only amplitudes to the denominator of the CP asymmetries is lengthy. Keeping only the terms in $ | \lambda_d |^2 $, $ | \lambda_s |^2 $, and $ \text{Re} \{ Q_{u d c s} \} $, i.e., neglecting $ | \lambda_b |^2 $, $ \text{Re} \{ Q_{u d c b} \} $, and $ \text{Re} \{ Q_{u s c b} \} $ (or, alternatively, neglecting contributions from penguin operators), we have:
\begin{eqnarray}\label{eq:BRs_CPasym_denominator}
\mathrm{den} \, (A_{CP}^i)_{I=0} &\!\!\! = &\!\!\! 2 \left( | \lambda_d |^2 | \Omega^{(0)}_{i 1} |^2 ( T^{C C}_{\pi \pi} )^2 + | \lambda_s |^2 | \Omega^{(0)}_{i 2} |^2 ( T^{C C}_{K K} )^2 + 2 \text{Re} \{ Q_{u d c s} \} \omega^{({\rm Re})}_i T^{C C}_{\pi \pi} T^{C C}_{K K} \right) 
\nonumber\\
	&\!\!\! \approx &\!\!\! 2 | \lambda_d |^2 \left( | \Omega^{(0)}_{i 1} |^2 ( T^{C C}_{\pi \pi} )^2 + | \Omega^{(0)}_{i 2} |^2 ( T^{C C}_{K K} )^2 - 2 \omega^{({\rm Re})}_i T^{C C}_{\pi \pi} T^{C C}_{K K} \right) , 
\end{eqnarray}

\noindent
where $ \omega^{({\rm Re})}_i \equiv \text{Re} \{ \Omega^{(0) \ast}_{i 1} \Omega^{(0)}_{i 2} \}$;
in what will follow, $ \omega^{({\rm Re})}_1 \equiv \omega^{({\rm Re})}_\pi $ and $ \omega^{({\rm Re})}_2 \equiv \omega^{({\rm Re})}_K $. Numerically,
\begin{equation}
    J / | \lambda_d |^2\, =\, r_{\rm CKM} \,\simeq\, 6.2 \times 10^{-4}\, ,
\end{equation}
so the numerator is typically much smaller than the denominator.

The previous exercise can be easily extended to isospin-two ($\pi\pi$) and isospin-one ($KK$) contributions, which we assume to be elastic. 
Although these are single-channel amplitudes,
they can also lead to contributions to the CP asymmetries when interfering with the corresponding isospin-zero contributions. Adopting the parameterization $ \Omega^{(1,2)} = |\Omega^{(1,2)}| \, e^{i \phi_{1,2}} $
(these quantities will later be extracted from branching ratios), one derives similar expressions in terms of rephasing-invariant quantities. The combinations analogous to $ \omega_i^{({\rm Im})} $ above are now:
%
\begin{eqnarray}
	\frac{\tilde{\omega}^{({\rm Im})}_{\pi i}}{|\Omega^{(2)}|} &\!\!\equiv &\!\! \text{Im} \{ \Omega^{(0)}_{1 i} e^{-i \phi_2} \} 
 \,, \\
    \frac{\tilde{\omega}^{({\rm Im})}_K}{|\Omega^{(1)}|} &\!\!\equiv &\!\! \text{Im} \{ \Omega^{(0) \ast}_{2 1} e^{i \phi_1} \}\, , 
\end{eqnarray}
where $ \phi_2 $ ($ \phi_1 $) is the strong phase developed by the isospin-two (respectively, isospin-one) amplitude.
Appearing in the branching ratios, we have the following extra quantities:
\begin{eqnarray}
    \frac{\tilde{\omega}^{({\rm Re})}_{\pi i}}{|\Omega^{(2)}|} &\!\!\equiv &\!\! \text{Re} \{ \Omega^{(0)}_{1 i} e^{-i \phi_2} \} 
\,, \\
    \frac{\tilde{\omega}^{({\rm Re})}_{K i}}{|\Omega^{(1)}|} &\!\!\equiv &\!\! \text{Re} \{ \Omega^{(0) \ast}_{2 i} e^{i \phi_1} \}\, .
\end{eqnarray}

\subsection{Rescattering parameters}

\begin{table}[t]
    \centering
    \renewcommand{\arraystretch}{1.2}
	\begin{tabular}{|c|c|c|c|}
		\hline
		$ A_{CP} ( \pi^- \pi^+) $; & \multirow{2}{*}{interference} & \multirow{2}{*}{expression} & final \\
        $ A_{CP} ( \pi^0 \pi^0) $ & & & numerics \\
		\hline
         & I=0/I=0 & $ 0.0019 \times \omega^{({\rm Im})}_\pi $ & 0.00027 \\
        \cline{2-4}
        numerator & \multirow{2}{*}{I=0/I=2} & $ 0.00041 \times \tilde{\omega}^{({\rm Im})}_{\pi 2} + 0.00026 \times \tilde{\omega}^{({\rm Im})}_{\pi 1} $; & -0.00009; \\
         & & $ -0.00081 \times \tilde{\omega}^{({\rm Im})}_{\pi 2} - 0.00052 \times \tilde{\omega}^{({\rm Im})}_{\pi 1} $ & 0.00018 \\
		\hline
         & I=0/I=0 & $ | \Omega^{(0)}_{1 1} |^2 + 0.57 \times | \Omega^{(0)}_{1 2} |^2 - 1.51 \times \omega^{({\rm Re})}_\pi $ & 1.11 \\
        \cline{2-4}
		\multirow{2}{*}{denominator} & \multirow{2}{*}{I=0/I=2} & $ 0.64 \times \tilde{\omega}^{({\rm Re})}_{\pi 1} - 0.49 \times \tilde{\omega}^{({\rm Re})}_{\pi 2} $; & 0.03; \\
         & & $ -1.28 \times \tilde{\omega}^{({\rm Re})}_{\pi 1} + 0.97 \times \tilde{\omega}^{({\rm Re})}_{\pi 2} $ & -0.07 \\
        \cline{2-4}
         & I=2/I=2 & $ |\Omega^{(2)}|^2 \times 0.10 $; $ |\Omega^{(2)}|^2 \times 0.41 $ & 0.08; 0.33 \\
		\hline
		\hline
		$ A_{CP} ( K^- K^+) $; & \multirow{2}{*}{interference} & \multirow{2}{*}{expression} & final \\
        $ A_{CP} ( K_S K_S) $ & & & numerics \\
		\hline
         & I=0/I=0 & $ 0.0019 \times \omega^{({\rm Im})}_K $ & -0.00032 \\
        \cline{2-4}
        numerator & \multirow{2}{*}{I=0/I=1} & $ 0.0019 \times \tilde{\omega}^{({\rm Im})}_K $; & -0.00019; \\
         & & $ -0.0019 \times \tilde{\omega}^{({\rm Im})}_K $ & 0.00019 \\
		\hline
        \multirow{4}{*}{denominator} & I=0/I=0 & $ | \Omega^{(0)}_{2 1} |^2 + 0.57 \times | \Omega^{(0)}_{2 2} |^2 - 1.51 \times \omega^{({\rm Re})}_K $ & 1.05 \\
        \cline{2-4}
         & \multirow{2}{*}{I=0/I=1} & $ 1.15 \times \tilde{\omega}^{({\rm Re})}_{K 2} - 1.51 \times \tilde{\omega}^{({\rm Re})}_{K 1} $; & 1.23; \\
         & & $ -1.15 \times \tilde{\omega}^{({\rm Re})}_{K 2} + 1.51 \times \tilde{\omega}^{({\rm Re})}_{K 1} $ & -1.23 \\
        \cline{2-4}
         & I=1/I=1 & $ |\Omega^{(1)}|^2 \times 0.57 $ & 0.36 \\
		\hline
	\end{tabular}
	\caption{Budget of contributions to the CP asymmetries. The column ``final numerics'' corresponds to the values found at Eq.~\eqref{eq:values_for_final_numerics}. When two values are provided, the first corresponds to the charged channels ($ D^0 \to \pi^- \pi^+ $ and $ D^0 \to K^- K^+ $), while the second to the neutral ones ($ D^0 \to \pi^0 \pi^0 $ and $ D^0 \to K_S K_S $). For the CP asymmetries of each channel, divide the sum of the corresponding `numerator' terms by the sum of the `denominator' ones.}\label{tab:budget}
\end{table}

Following the previous discussion,
we have the following 17 parameters describing rescattering effects:
\begin{eqnarray}\label{eq:rescattering_parameters}
    && \omega^{({\rm Im})}_\pi \,, \omega^{({\rm Im})}_K \,, \tilde{\omega}^{({\rm Im})}_{\pi 1} \,, \tilde{\omega}^{({\rm Im})}_{\pi 2} \,, \tilde{\omega}^{({\rm Im})}_K \,, \\
    && \tilde{\omega}^{({\rm Re})}_{\pi 1} \,, \tilde{\omega}^{({\rm Re})}_{\pi 2} \,, \tilde{\omega}^{({\rm Re})}_{K 1} \,, \tilde{\omega}^{({\rm Re})}_{K 2} \,, | \Omega^{(0)}_{1 1} |^2 \,, | \Omega^{(0)}_{1 2} |^2 \,, | \Omega^{(0)}_{2 1} |^2 \,, | \Omega^{(0)}_{2 2} |^2 \,, \omega^{({\rm Re})}_\pi \,, \omega^{({\rm Re})}_K \,, |\Omega^{(1)}| \,, |\Omega^{(2)}| \, , \nonumber
\end{eqnarray}
which are functions of the 12 parameters 
$\mathrm{Re}\, \{ \Omega^{(0)}_{i j} \}$, $\mathrm{Im}\, \{ \Omega^{(0)}_{i j} \}$,
$|\Omega^{(1,2)}|$, $ \phi_{1, 2} $, $i, j = 1, 2$.
The parameters $ |\Omega^{(1)}| $ and $ |\Omega^{(2)}| $ can be directly extracted from the branching ratios $ D^+ \to K_S K^+ $ and $ D^+ \to \pi^0 \pi^+ $, respectively. This results in:\footnote{Hereafter, the Wilson coefficients and quark masses are taken at $2$~GeV.}
%
\begin{equation}\label{eq:f1_f2}
    |\Omega^{(1)}| = 0.79 \,, \qquad \qquad |\Omega^{(2)}| = 0.90 \, .
\end{equation}
There are further four branching ratios of $ D^0 $ decays, that depend linearly on 10 parameters, namely, $ | \Omega^{(0)}_{1 1} |^2 $, $ | \Omega^{(0)}_{1 2} |^2 $, $ | \Omega^{(0)}_{2 1} |^2 $, $ | \Omega^{(0)}_{2 2} |^2 $, $ \tilde{\omega}^{({\rm Re})}_{\pi 1} $, $ \tilde{\omega}^{({\rm Re})}_{\pi 2} $, $ \tilde{\omega}^{({\rm Re})}_{K 1} $, $ \tilde{\omega}^{({\rm Re})}_{K 2}$, $ \omega^{({\rm Re})}_\pi $, $ \omega^{({\rm Re})}_K $
(that depend on the 10 quantities $\mathrm{Re}\, \{ \Omega^{(0)}_{i j} \}$, $\mathrm{Im}\, \{ \Omega^{(0)}_{i j} \}$ and $ \phi_{1, 2} $).
Therefore, by using only the branching ratios, the set of these parameters remains under-constrained.

However, the numerators of the CP asymmetries only depend on the 5 remaining parameters, namely, $ \omega^{({\rm Im})}_\pi $, $ \omega^{({\rm Im})}_K $, $ \tilde{\omega}^{({\rm Im})}_{\pi 1} $, $ \tilde{\omega}^{({\rm Im})}_{\pi 2} $, $ \tilde{\omega}^{({\rm Im})}_K $.
Fixing the denominators of the CP asymmetries, which are proportional to the branching ratios, to their experimental values we have then that the four CP asymmetries of the $ D^0 \to \pi^- \pi^+, \pi^0 \pi^0, K^- K^+, K_S K_S $ modes depend linearly on 5 parameters. Using the measurements by LHCb \cite{LHCb:2019hro,LHCb:2022vcc} is not enough then to determine ranges for the remaining two CP asymmetries in the final modes containing neutral pions and kaons.
In a companion paper \cite{PSVS:2023}, we discuss how the use of the determinant of the Omn\`{e}s matrix, which has the great advantage of being independent of the inelasticity, helps in setting ranges for the rescattering parameters controlling the level of CP asymmetry. Moreover, as discussed therein, the use of Eq.~\eqref{eq:main_discontinuity_eq} leads to an additional relation, namely
\begin{equation}
    {\rm Im} \{ \Omega^{(0) \dagger} (s) \, \Sigma \, \Omega^{(0)} (s) \} = 0 \;\; \Rightarrow \;\; \sigma_\pi \, \omega^{({\rm Im})}_\pi + \sigma_K \, \omega^{({\rm Im})}_K = 0
\end{equation}
that implies that $\omega^{({\rm Im})}_\pi$ and $\omega^{({\rm Im})}_K$ have opposite signs, and similar absolute values, thus reducing the number of parameters controlling the CP asymmetries to 4.

The dependence of the CP asymmetries on the rescattering parameters is illustrated in the previous to the last column of Tab.~\ref{tab:budget}.
Note that the interference terms I=0/I=2, I=2/I=2, I=0/I=1, and I=1/I=1 are sources of difference among pion and kaon channels independently of the rescattering parameters. On the other hand, the interference terms I=0/I=0 for pions and kaons have the same pre-factors, see Eqs.~\eqref{eq:CPasym_numerator} and \eqref{eq:BRs_CPasym_denominator}, and the difference comes from the rescattering parameters, namely, $ |\omega^{({\rm Im})}_\pi| \neq |\omega^{({\rm Im})}_K| $, $ | \Omega^{(0)}_{1 1} |^2 \neq | \Omega^{(0)}_{2 1} |^2 $, $ | \Omega^{(0)}_{1 2} |^2 \neq | \Omega^{(0)}_{2 2} |^2 $, $ |\omega^{({\rm Re})}_\pi| \neq |\omega^{({\rm Re})}_K| $.

In the following section, except for $ |\Omega^{(1)}| $ and $ |\Omega^{(2)}| $, for which we consider Eq.~\eqref{eq:f1_f2}, the remaining rescattering parameters in Eq.~\eqref{eq:rescattering_parameters} are extracted from the use of DRs.

\subsection{Results based on DRs}




\begin{table}[h!]
    \centering
    \renewcommand{\arraystretch}{1.6}
    \begin{tabular}{|c|c|}
        \hline
        $ \frac{\mathcal{B} ( D^0 \to \pi^- \pi^+)_{theo;CV}}{\mathcal{B} ( D^0 \to \pi^- \pi^+)_{exp;CV}} $ & $ 1.1 $ \\
        $ \frac{\mathcal{B} ( D^0 \to \pi^0 \pi^0)_{theo;CV}}{\mathcal{B} ( D^0 \to \pi^0 \pi^0)_{exp;CV}} $ & $ 1.1 $ \\
        $ \frac{\mathcal{B} ( D^+ \to \pi^0 \pi^+)_{theo;CV}}{\mathcal{B} ( D^+ \to \pi^0 \pi^+)_{exp;CV}} $ & fixed to 1 \\
        \hline
        $ A_{CP} ( D^0 \to \pi^- \pi^+) \times 10^4 $ & $ 2 $; $ 3 $ \\
        $ A_{CP} ( D^0 \to \pi^0 \pi^0) \times 10^4 $ & $ 3 $; $ 0.5 $ \\
        $ A_{CP} ( D^+ \to \pi^0 \pi^+) $ & $ 0 $ \\
        \hline
        $ | \mathcal{A}^\pi_2 | \times 10^6 $ & $ 0.5 \times |\Omega^{(2)}| $ \\
        $ | \mathcal{A}^\pi_0 | \times 10^6 $ & $ 1.2 $ \\
        $ | \mathcal{B}^\pi_2 | / r_{\rm CKM} \times 10^6 $ & $ 0.5 \times |\Omega^{(2)}| $ \\
        $ | \mathcal{B}^\pi_0 | / r_{\rm CKM} \times 10^6 $ & $ 0.8 $ \\
        $ \arg(\mathcal{A}^\pi_0) $ & $ 93^{\rm o} $ \\
        $ \arg(\mathcal{B}^\pi_0) $ & $ -72^{\rm o} $ \\
        \hline
        \multicolumn{1}{c}{} & \multicolumn{1}{c}{} \\
    \end{tabular}
    \begin{tabular}{|c|c|}
        \hline
        $ \frac{\mathcal{B} ( D^0 \to K^- K^+)_{theo;CV}}{\mathcal{B} ( D^0 \to K^- K^+)_{exp;CV}} $ & $ 1.1 $ \\
        $ \frac{\mathcal{B} ( D^0 \to K_S K_S)_{theo;CV}}{\mathcal{B} ( D^0 \to K_S K_S)_{exp;CV}} $ & $ 1.1 $ \\
        $ \frac{\mathcal{B} ( D^+ \to K_S K^+)_{theo;CV}}{\mathcal{B} ( D^+ \to K_S K^+)_{exp;CV}} $ & fixed to 1 \\
        \hline
		$ A_{CP} ( D^0 \to K^- K^+) \times 10^{4} $ & $ -2 $ \\
		$ A_{CP} ( D^0 \to K_S K_S) \times 10^{4} $ & $ -7 $ \\
		$ A_{CP} ( D^+ \to K_S K^+) $ & $ 0 $ \\
		\hline
		$ | \mathcal{A}^K_{11} | \times 10^6 $ & $ 0.8 \times |\Omega^{(1)}| $ \\
		$ | \mathcal{A}^K_0 | \times 10^6 $ & $ 1.1 $ \\
		$ | \mathcal{B}^K_{11} | / r_{\rm CKM} \times 10^6 $ & $ 0.3 \times |\Omega^{(1)}| $ \\
		$ | \mathcal{B}^K_0 | / r_{\rm CKM} \times 10^6 $ & $ 0.9 $ \\
		$ \arg(\mathcal{A}^K_0) $ & $ -66^{\rm o} $ \\
		$ \arg(\mathcal{B}^K_0) $ & $ 95^{\rm o} $ \\
		$ | \mathcal{A}^K_{13} |, | \mathcal{B}^K_{13} | $ & sub-leading $ \frac{1}{N_C} $ \\
        \hline
    \end{tabular}
	\caption{Predictions based on the reference solution of Tab.~\ref{tab:num_sols_Omnes}. The notation $\mathcal{A}$ ($\mathcal{B}$) designates CP-even (respectively, CP-odd) amplitudes; ``CV'' stands for central value. When two numerical values are provided, the first corresponds to $ \phi_2 \simeq 0 $, while the second to $ \phi_2 \simeq \pm \pi $.}\label{tab:full_set_predictions}
\end{table}

Before discussing predictions for CP asymmetries, we need to ensure that branching ratios can be correctly reproduced.
Rescattering effects in isospin-zero are given in Tab.~\ref{tab:num_sols_Omnes} for various situations.\footnote{For illustrative purposes only, the procedure of Refs.~\cite{Bauer:1986bm,Smith:1998nu} leads to ($ S_S^{1/2} = \pm O \, \sigma \, D^{1/2} \, O^T $ if $ S_S = O \, D \, O^T $, where $O$ is an orthogonal matrix, $D$ is a diagonal matrix, and $\sigma$ is another diagonal matrix with $\pm 1$ elements):

\begin{equation}
    S_S^{1/2} (M_D^2) = \pm \begin{pmatrix}
        0.68 \, e^{-0.61 \, i} & 0.74 \, e^{+1.05 \, i} \\
        0.74 \, e^{+1.05 \, i} & 0.68 \, e^{-0.44 \, i} \\
    \end{pmatrix} \,, \quad \text{or} \quad
    S_S^{1/2} (M_D^2) = \pm \begin{pmatrix}
        0.74 \, e^{-2.02 \, i} & 0.67 \, e^{+2.62 \, i} \\
        0.67 \, e^{+2.62 \, i} & 0.74 \, e^{-2.17 \, i} \\
    \end{pmatrix}
\end{equation}
when using the same inputs used to generate the reference solution.}
We find that Omn\`{e}s solutions resulting from sols.~II and III, and sols.~B' and C' do not lead to branching ratios of charm-meson decays in agreement with their experimental values, simultaneously for all four $ D^0 \to \pi^- \pi^+ $, $ \pi^0 \pi^0 $, $ K^- K^+ $, and $ K_S K_S $ transitions. However, we highlight that a set of solutions is found satisfying the latter constraint, resulting from sol.~I for the phase-shift $ \delta^0_0 $ and inelasticity, and given in the first column of Tab.~\ref{tab:num_sols_Omnes}. As previously stated, the profile of the inelasticity carries a large uncertainty, and solutions leading to the correct branching ratios
are found when varying the inelasticity inside its error bar towards smaller values (i.e., away from the elastic limit), referred to as $ \eta^0_0 - \delta \eta^0_0 $.
We display in Tab.~\ref{tab:num_sols_Omnes} three such solutions, that differ in the way the asymptotic value of the inelasticity is approached, corresponding to different values of $ m^\ast_\eta $. In what follows, the reference case refers to $ m^\ast_\eta = 2 $, although anyways $ m^\ast_\eta = 1 $ or $ m^\ast_\eta = 3 $ lead to similar Omn\`{e}s solutions.

\begin{figure}[t]
	\centering
    \includegraphics[scale=0.38]{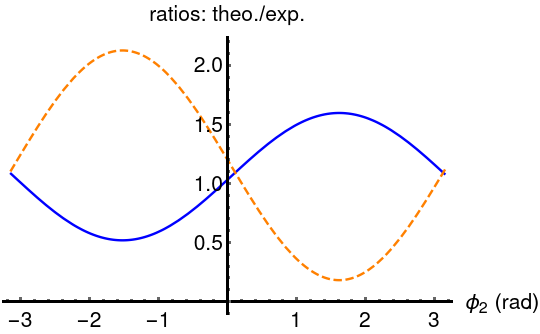} \hspace{2mm}
    \includegraphics[scale=0.38]{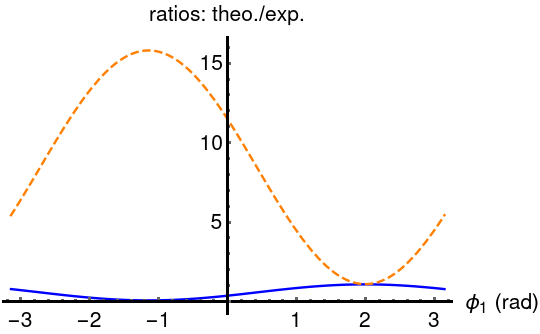} \\
    \vspace{3mm}
    \includegraphics[scale=0.38]{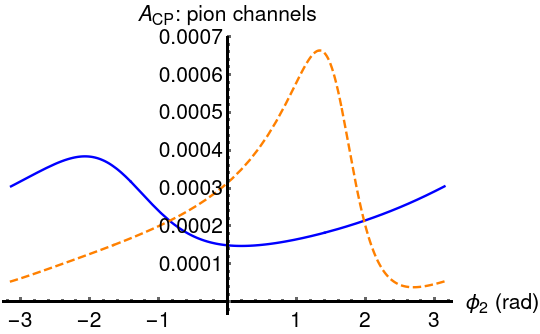} \hspace{2mm}
    \includegraphics[scale=0.38]{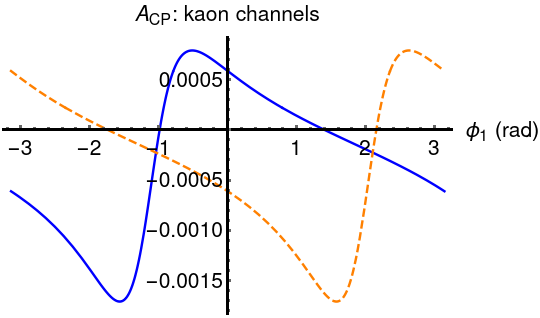}
	\caption{Physical predictions for the reference case of Tab.~\ref{tab:num_sols_Omnes}. Charged modes are shown in solid blue, while neutral ones are shown in dashed orange.
	Left (right) panels correspond to pion (kaon) modes. The top panels show the ratio of the theoretical and experimental $D^0\to P^-P^+$ branching ratios, as function of the relevant $\phi_i$ phases, while the lower panels display the corresponding CP asymmetries.}
    \label{fig:predictions}
\end{figure}

Having selected the Omn\`{e}s solutions based on the branching ratios, we then predict the CP asymmetries.
In Tab.~\ref{tab:budget} we give numerical details about the predictions of CP asymmetries in charm-meson decays.
Observables are illustrated in Fig.~\ref{fig:predictions}. Two cases of the phase-shift $ \phi_2 $ for isospin-two lead to the correct branching ratios simultaneously for $ \pi^0 \pi^0 $ and $ \pi^- \pi^+ $, namely, $ \phi_2 \simeq \pm \pi $, and $ \phi_2 \simeq 0 $, which is closer to Ref.~\cite{Durusoy:1973aj} and should therefore be preferred.
In the reference case of Tab.~\ref{tab:num_sols_Omnes}:

\begin{eqnarray}\label{eq:values_for_final_numerics}
    && \omega^{({\rm Im})}_\pi = 0.15 \,, \; \tilde{\omega}^{({\rm Im})}_{\pi 1} = 0.53 \,, \; \tilde{\omega}^{({\rm Im})}_{\pi 2} = -0.57 \,, \; \omega^{({\rm Im})}_K = -0.17 \,, \; \tilde{\omega}^{({\rm Im})}_K = -0.1 \,, \nonumber\\
    && | \Omega^{(0)}_{1 1} |^2 = 0.34 \,, \; | \Omega^{(0)}_{1 2} |^2 = 0.42 \,, \; \omega^{({\rm Re})}_\pi = -0.35 \,, \nonumber\\
    && | \Omega^{(0)}_{2 1} |^2 = 0.35 \,, \; | \Omega^{(0)}_{2 2} |^2 = 0.38 \,, \; \omega^{({\rm Re})}_K = -0.32 \,, \nonumber\\
    && \tilde{\omega}^{({\rm Re})}_{\pi 1} = -0.07 \,, \; \tilde{\omega}^{({\rm Re})}_{\pi 2} = -0.16 \,, \; \tilde{\omega}^{({\rm Re})}_{K 1} = -0.45 \,, \; \tilde{\omega}^{({\rm Re})}_{K 2} = 0.47\, .
\end{eqnarray}
These values correspond to $ \phi_2 \simeq 0 $, $ \phi_1 = 2.0 $. For $ \phi_2 = \pm \pi $, $\tilde{\omega}^{({\rm Im})}_{\pi 1}, \tilde{\omega}^{({\rm Im})}_{\pi 2}, \tilde{\omega}^{({\rm Re})}_{\pi 1}, \tilde{\omega}^{({\rm Re})}_{\pi 2} $ flip signs with respect to $ \phi_2 = 0 $.

In both cases of $ \phi_2 $, the main contribution to the CP asymmetry $ D^0 \to \pi^- \pi^+ $ ($ D^0 \to K^- K^+ $) comes from the interference term I=0/I=0 (as well, I=0/I=0), followed closely by I=0/I=2 (respectively, I=0/I=1).
For the I=0/I=2 contribution, we observe a cancellation due to the fact that $ \tilde{\omega}^{({\rm Im})}_{\pi 1} $ and $ \tilde{\omega}^{({\rm Im})}_{\pi 2} $ carry opposite signs.
We obtain that the predicted values of the CP asymmetries are too small in the charged decay modes to reproduce the measured value of $ \Delta A_{CP}^{\mathrm{dir}} $ \cite{LHCb:2019hro}.
In the case of $ \phi_2 \simeq 0 $, the two interference terms I=0/I=2 and I=0/I=0 contributing to $ A_{CP} ( D^0 \to \pi^- \pi^+) $ largely cancel, but they add up in the case $ \phi_2 \simeq \pm \pi $.
However, were there no cancellations (i.e., by artificially reversing signs to obtain a constructive pattern), the level of CP violation would remain small compared to the experimental measurement by LHCb.
The value of the CP asymmetry for $ D^0 \to K_S K_S $ is potentially large, at the price of a small branching ratio, see App.~\ref{app:num_inputs}.

As previously noticed, rescattering parameters are a source of breaking of a potential symmetry relating charm-meson decays into pion and kaon pairs: $ |\omega^{({\rm Im})}_\pi| \neq |\omega^{({\rm Im})}_K| $ at the level of 20\%, and $ | \Omega^{(0)}_{1 2} |^2 \neq | \Omega^{(0)}_{2 2} |^2 $ and $ |\omega^{({\rm Re})}_\pi| \neq |\omega^{({\rm Re})}_K| $ at the level of 10\%, while $ | \Omega^{(0)}_{1 1} |^2 \simeq | \Omega^{(0)}_{2 1} |^2 $. This breaking between isospin-zero amplitudes should be compared to the level of $SU(3)_F$ breaking found in decay constants and form factors, at the level of 20\%, see App.~\ref{app:num_inputs}.


Further numerical information is provided in Tab.~\ref{tab:full_set_predictions}.
Note that rescattering effects lead to different strong phases for the isospin-zero amplitudes $ \mathcal{A}^\pi_0 $ with respect to $ \mathcal{B}^\pi_0 $, and also $ \mathcal{A}^K_0 $ with respect to $ \mathcal{B}^K_0 $.
When $ \Omega^{(0)}_{1 2} = 0 $,
\begin{equation}
    \tan{(\arg \mathcal{A}^\pi_0)} = \text{Im} [ \Omega^{(0)}_{1 1} ] / \text{Re} [ \Omega^{(0)}_{1 1} ] = \tan{(\arg \mathcal{B}^\pi_0)}\, .
\end{equation}
Also, when $ \Omega^{(0)}_{2 1} = 0 $,
\begin{equation}
    \tan{(\arg \mathcal{A}^K_0)} = \text{Im} [ \Omega^{(0)}_{2 2} ] / \text{Re} [ \Omega^{(0)}_{2 2} ] = \tan{(\arg \mathcal{B}^K_0)} \,.
\end{equation}
Having instead $ \Omega^{(0)}_{1 2} \neq 0 $ and/or $ \Omega^{(0)}_{2 1} \neq 0 $ allows then for contributions to the CP asymmetries coming from the interference term I=0/I=0.

The numerical conclusions made above do not depend significantly on the scale used for the Wilson coefficients and quark masses, which has been taken at 2~GeV in Eq.~\eqref{eq:f1_f2} and Tabs.~\ref{tab:budget} and \ref{tab:full_set_predictions}.

We stress that the work of a companion paper circumvents the need to discuss the input for the inelasticity \cite{PSVS:2023}, which carries a large uncertainty, and one achieves bounds on the CP asymmetries rather than predictions as above.

\section{Conclusions}\label{sec:conclusions}



CP violation has been recently established in the charm sector, and its prediction based on the SM represents an outstanding problem due to the presence of non-perturbative QCD effects.
In charm physics, the mechanism of CP violation is expected to be largely influenced by such long-distance effects,
while short-distance penguin contributions are expected to play a less important role.
It is essential then to include rescattering effects in order to build an SM prediction of the recently measured CP asymmetries.

We have discussed a data-driven approach, which is based on the use of dispersion relations to take into account rescattering in the isospin-zero mode, with the subtraction constants being given by large $N_C$.
Only pion and kaon pairs are included in the analysis, and further inelasticities are omitted. 
Given the large uncertainties attached to the pion-kaon inelasticity,
we use $D^0 \to \pi^- \pi^+, \pi^0 \pi^0, K^- K^+, K_S K_S$ branching ratios to limit this source of hadronic uncertainties.
We have also employed the charged decay modes $ D^+ \to K_S K^+ $ and $ D^+ \to \pi^0 \pi^+ $ to extract rescattering quantities for isospin-one and -two, respectively.
There are four non-perturbative quantities controlling the CP asymmetries, that are determined by the dispersion relations (a companion paper discusses bounds on these quantities).
Our main result is that CP asymmetries in the $D^0\to\pi^-\pi^+$ and $D^0 \to K^-K^+$ decay modes are too small compared to the experimental value \cite{LHCb:2019hro}.
The main reason for this is not an accidental cancellation among contributions, but rather that rescattering effects turn out not producing enough enhancement.
We also find that the level of $SU(3)_F$ breaking due to rescattering effects in isospin-zero amplitudes is similar to the one of decay constants and form factors.

In the future, we also plan to address further inelasticities. Their effect might be expected not to be too large though: in the cases of $ \rho $ pairs and $a_1 (1260) \pi$, whose thresholds take place respectively at $ \sim 1.54 $~GeV and $ \sim 1.23 $~GeV, there is a phase-space suppression.
Decay modes with $ \eta^{(')} $ are expected to give small contributions.
In any case, if such effects are important this means that a similar level of CP violation already found experimentally in $D^0 \to \pi^- \pi^+, K^- K^+$ should also be found in other charm-meson decay channels.




\section*{Acknowledgements}

We would like to thank specially Bachir~Moussallam for many crucial discussions for the development of this project, in particular concerning fundamental solutions of Ref.~\cite{Muskhelishvili}, the numeric method of Ref.~\cite{Moussallam:1999aq} and off-diagonal $T$-matrix elements of Refs.~\cite{Buettiker:2003pp,Garcia-Martin:2010kyn}; also, Arkaitz~Rodas for kindly providing further details and numerical files about Refs.~\cite{Pelaez:2019eqa,Pelaez:2020gnd};
finally, Miguel~Albaladejo, V\'{e}ronique~Bernard, Joachim~Brod, S\'{e}bastien~Descotes-Genon, Hector~Gisbert~Mullor, Sebastian~J\"{a}ger, Martin~Jung, Patr\'{i}cia~C.~Magalh\~{a}es, Ulrich~Nierste, Emilie~Passemar, Jos\'{e}~R. Pel\'{a}ez, Marcos~N.~Rabelo, and Jaume~Tarr\'{u}s~Castell\`{a} for useful discussions.

\vspace{3mm}
\noindent
This work has been supported by MCIN/AEI/10.13039/501100011033, grants PID2020-114473GB-I00 and PRE2018-085325, and by Generalitat Valenciana, grant PROMETEO/2021/071.
This project has received funding from the European Union’s Horizon 2020 research and innovation programme under the Marie Sklodowska-Curie grant agreement No 101031558.
LVS is grateful for the hospitality of the CERN-TH group where part of this research was executed.

\appendix

\section{Numerical inputs}\label{app:num_inputs}



The Wilson coefficients $ C_1, \ldots, C_6 $ are given in Tab.~\ref{tab:WCs}, based in Ref.~\cite{deBoer:2016dcg}, at NLO
in the naive dimensional regularization (NDR) scheme; one observes at this order a strong scheme dependence (NDR vs. the `t Hooft-Veltman scheme), see Ref.~\cite{Buchalla:1995vs}.

\begin{table}[t]
	\centering
	\begin{tabular}{|c|c|c|c|c|c|c|}
	\hline
	$ \mu $ & $ C_1 $ & $ C_2 $ & $ C_3 $ & $ C_4 $ & $ C_5 $ & $ C_6 $ \\ 
	\hline
	\hline
	$ m_c $ & $ 1.22 $ & $ -0.40 $ & $ 0.021 $ & $ -0.055 $ & $ 0.0088 $ & $ -0.060 $ \\
	\hline
	$ 2 $~GeV & $ 1.18 $ & $ -0.32 $ & $ 0.011 $ & $ -0.031 $ & $ 0.0068 $ & $ -0.032 $ \\
	\hline
	\end{tabular}
	\\
	\vspace{0.5cm}
	\begin{tabular}{|c|c|c|c|c|c|}
	\hline
	$ \mu $ & $ m_u $ & $ m_d $ & $ \hat{m} \equiv (m_u + m_d)/2 $ & $ m_s $ & $ m_c $ \\
	\hline\hline
	$ m_c $ & $2.50 \pm 0.09$ & $5.48 \pm 0.06$ & $ 4.00 \pm 0.06 $ & $109.0 \pm 0.7$ & $1280 \pm 13$ \\
	\hline
    $2$~GeV & $2.14 \pm 0.08$ & $4.70 \pm 0.05$ & $ 3.427 \pm 0.051 $ & $93.46 \pm 0.58$ & $1097 \pm 11$ \\
	\hline
	\end{tabular}
	\caption{In the upper panel, the Wilson coefficients at NLO in the NDR scheme, with four dynamical flavours, see \cite{deBoer:2016dcg} and references therein; $ \alpha_s (M_Z) = 0.1179 $ (we employ its expression at NLO), $ \mu_b = m_b $, with $m_b = 4.18$~GeV, and $M_W = 80.4$~GeV, $M_Z = 91.1876$~GeV, $m_t = 163.3$~GeV. The bottom panel gives the $ \overline{\text{MS}} $ quark masses in MeV at $ N_f = 2+1+1 $, see \cite{FlavourLatticeAveragingGroupFLAG:2021npn} and references therein; the running factor 0.857 for $m_c$ to $2$~GeV has been employed.
	}\label{tab:WCs}
\end{table}

The following values of the form factors and decay constants, obtained from lattice simulations with $ N_f = 2+1+1 $ active quark flavours, are taken from Ref.~\cite{FlavourLatticeAveragingGroupFLAG:2021npn}, see also references therein:
%
\begin{eqnarray}
    && \frac{f_K}{f_\pi} = 1.1934 \pm 0.0019 \,, \qquad f_K = (0.1557 \pm 0.0003)~\rm{GeV} \,, \\
    && f_D = (0.2120 \pm 0.0007)~\rm{GeV} \,, \\
    && f_0^{D \pi} (0) = 0.612 \pm 0.035 \,, \qquad f_0^{D K} (0) = 0.7385 \pm 0.0044\, .
\end{eqnarray}
We consider the following single-pole corrections to the form factors \cite{Muller:2015lua}, which amount to a tiny correction
\begin{equation}
	f_0^{D \pi} (M_\pi^2) = \frac{f_0^{D \pi} (0)}{1 - \frac{M_\pi^2}{M^2_{D_0^\ast} (2300)}} \,, \qquad\qquad
 f_0^{D K} (M_K^2) = \frac{f_0^{D K} (0)}{1 - \frac{M_K^2}{M^2_{D_{s 0}^\ast}\, . (2317)^\pm}} \,.
\end{equation}
\noindent
For the meson masses we adopt the values:
$ M_\pi = 139.57 $~MeV,
$ M_K = 496 $~MeV,
$ M_D = 1864.84 $~MeV,
$ M_{D_0^\ast} (2300) = (2343 \pm 10) $~MeV,
$ M_{D_{s 0}^\ast} (2317)^\pm = (2317.8 \pm 0.5) $~MeV;
$ D^{0, \pm} $ lifetimes are $ \tau_{D^\pm} = 1.033 $~ps, and $ \tau_{D^0} = 0.4103 $~ps \cite{Workman:2022ynf}.

The entries of the CKM matrix are taken from the CKMfitter Spring `21 \cite{Charles:2004jd,CKMfitter}
values of the Wolfenstein parameters:
\begin{eqnarray}
	&& A = 0.8132 \,,\quad \lambda = 0.22500 \,,\quad \bar{\rho} = 0.1566 \,,\quad \bar{\eta} = 0.3475 \,, \\
	&& \text{Re} \{ \lambda_d \} = -0.22 \,, \qquad \text{Im} \{ \lambda_d \} = 1.3 \times 10^{-4} \,, \\
	&& \text{Re} \{ \lambda_s \} = 0.22 \,, \qquad \text{Im} \{ \lambda_s \} = 6.9 \times 10^{-6} \,, \\
	&& \text{Re} \{ \lambda_b \} = 6.1 \times 10^{-5} \,, \qquad \text{Im} \{ \lambda_b \} = -1.4 \times 10^{-4}\, .
\end{eqnarray}

The relevant branching ratios have the following numerical values \cite{HFLAV:2022pwe}:
\begin{eqnarray}
    && \mathcal{B} ( D^0 \to K^- \pi^+) = ( 3.999 \pm 0.006 \pm 0.031 \pm 0.032 ) \, \% \,, \\
    && \mathcal{B} ( D^0 \to \pi^- \pi^+) = ( 0.1490 \pm 0.0012 \pm 0.0015 \pm 0.0019 ) \, \% \,, \\
    && \mathcal{B} ( D^0 \to K^- K^+) = ( 0.4113 \pm 0.0017 \pm 0.0041 \pm 0.0025 ) \, \%\, ,
\end{eqnarray}
with a correlation matrix
\begin{equation}
    corr( \mathcal{B} (K^- \pi^+), \mathcal{B} (\pi^- \pi^+), \mathcal{B} (K^- K^+) ) = \begin{pmatrix}
        1.00 & 0.77 & 0.76 \\
        0.77 & 1.00 & 0.58 \\
        0.76 & 0.58 & 1.00 \\
    \end{pmatrix}\, ,
\end{equation}
and \cite{Workman:2022ynf}
\begin{eqnarray}
    && \mathcal{B} ( D^0 \to \pi^0 \pi^0) = ( 0.826 \pm 0.025 ) \times 10^{-3} \,, \\
    && \mathcal{B} ( D^0 \to K_S K_S) = ( 0.141 \pm 0.005 ) \times 10^{-3} \,, \\
    && \mathcal{B} ( D^+ \to \pi^0 \pi^+) = ( 1.247 \pm 0.033 ) \times 10^{-3} \,, \\
    && \mathcal{B} ( D^+ \to K_S K^+) = ( 3.04 \pm 0.09 ) \times 10^{-3} \,, \\
    && \mathcal{B} ( D^+ \to K_L K^+) = ( 3.21 \pm 0.11 \pm 0.11 ) \times 10^{-3}\, .
\end{eqnarray}

In addition to the recent measurements in Eqs.~\eqref{eq:CP_discovery_in_charm} and (\ref{eq:ExpCPasym}), experimental values have been determined for the following CP asymmetries (combining direct and indirect CP violation in the case of $ D^0 $ decays) \cite{HFLAV:2022pwe}:
\begin{eqnarray}
    && A_{CP} ( D^0 \to \pi^0 \pi^0) = ( -0.03 \pm 0.64 ) \, \% \,, \\
    && A_{CP} ( D^0 \to K_S K_S) = ( -1.9 \pm 1.0 ) \, \% \,, \\
    && A_{CP} ( D^+ \to K_S K^+) = ( -0.11 \pm 0.25 ) \, \% \,, \\
    && A_{CP} ( D^+ \to (K^0 / \overline{K}^0) K^+) = ( +0.01 \pm 0.07 ) \, \% \,,
\end{eqnarray}
and \cite{Workman:2022ynf}:
\begin{equation}
    A_{CP} ( D^+ \to K_L K^+) = ( -4.2 \pm 3.2 \pm 1.2 ) \, \%\, .
\end{equation}

The inputs for phase-shifts and inelasticity have been discussed in Sec.~\ref{sec:inputs_DRs}.

\section{Numerical solution of the DRs}\label{app:num_sol_DRs}

\subsection{Numerical method}\label{app:num_sol_DRs_method}


We comment on the numerical method used to solve the DRs, which is based on the Legendre-Gauss quadrature \cite{Moussallam:1999aq,Jamin:2001zq} (an iteration strategy is followed by Refs.~\cite{Donoghue:1990xh,Celis:2013xja}).
Consider the following homogeneous problem:
%
%
%
\begin{equation}
	R(s) = \frac{1}{\pi} \dashint^\infty_{4 M^2} d s'\, \frac{1}{s' - s} X (s') R (s') \,, \quad X (s') = \tan \delta (s') \,, \quad R(s) = {\rm Re}\, (F(s))\, .
\end{equation}
%
We start by writing two basic properties of Legendre functions (in the interval $ -1 < z < 1 $, $ Q_j (z) $ is real; we take it real also outside this interval):\footnote{There was an unexpected difficulty, seemingly undocumented, when using \texttt{Python} 3.0 built-in functions \texttt{lqn} and \texttt{lqmn}, which do not return correct values for $ Q_j (u) $ for large negative $u$ and/or for $u \gtrsim -1$.}
%
\begin{equation}
	\dashint^1_{-1} d u\, \frac{P_{j} (u)}{u - z} = - 2 Q_{j} (z) \,, \qquad \int^1_{-1} d u\, P_{m} (u) P_{n} (u) = \delta_{m n} \frac{2}{2 m + 1}\, .
\end{equation}
We exploit this relation to write:
%
\begin{eqnarray}\label{eq:eqApprox}
	\dashint^1_{-1} d u'\, \frac{1}{u' - u} Y (u') &\approx &  - \sum^{N-1}_{j=0} (2 j + 1)\, Q_j (u) \int^1_{-1} d u'\, P_j (u') Y (u') \\
	&\approx & - \sum^{N-1}_{j=0} (2 j + 1)\, Q_j (u) \left[ \sum_{i=1}^M w_i P_j (u_i)\, Y (u_i) + R_M (P_j Y) \right] ,\nonumber
\end{eqnarray}
where in the first line we exploit the relation among Legendre polynomials of first and second degrees, and in the second line we execute a Gaussian quadrature, where the expressions for remainders in Gauss's formulas of quadrature integration are found in Ref.~\cite{Abramowitz} (Chapter 25.4):
%
\begin{equation}
	R_M (f) = \frac{2^{(2 M + 1)} (M !)^4}{(2 M + 1) [(2 M) !]^3} \;\frac{d^{(2 M)} f (x)}{d x^{(2 M)}} \Big|_{x = \xi} \qquad\qquad ( -1 < \xi <  1)\, .
\end{equation}
Therefore, if the remainder $ R_M (f) $ is sufficiently small,
%
\begin{equation}
	\dashint^b_a d s'\, \frac{1}{s' - s_k} X(s') R(s') \approx \sum^M_{i=1} \hat{W}_i \left[ 1 + \frac{2 (s_k - b)}{b - a} \right] X(s_i) R(s_i) \,, \quad\; s_i = \frac{a + b + (b - a) u_i}{2}\, ,
\end{equation}
\begin{equation}
	\dashint^\infty_a d s'\, \frac{1}{s' - s_k} X(s') R(s') \approx \sum^M_{i=1} \hat{W}_i \left[ 1 - \frac{2 a}{s_k} \right] \frac{s_i}{s_k} X(s_i) R(s_i) \,, \qquad s_i = \frac{2 a}{1 - u_i}\, ,
\end{equation}
%
\begin{equation}
	\hat{W}_i [z] = - w_i \sum^{N-1}_{j = 0} (2 j + 1)\, P_j (u_i)\, Q_j (z) \,, \qquad w_i = \frac{2}{1 - u_i^2} \left[ \frac{d P_M}{d u} (u_i) \right]^{-2} .
\end{equation}


In our case, we have subtractions and the system is inhomogeneous.
%
For $ n > 0 $ subtractions, choosing $ s_0 $ on the real axis below the cut $ s \ge 4 M^2 $,
\begin{equation}
	R(s) = \sum^{n-1}_{k=0} \frac{(s-s_0)^{k}}{k!} R^{(k)} (s_0) + \frac{(s - s_0)^n}{\pi} \dashint^\infty_{4 M^2} d s'\, \frac{1}{s' - s} X (s') \frac{R(s')}{(s' - s_0)^n}\, ,
\end{equation}
with $R^{(k)}$ the $k$th derivative,
for which a similar discussion holds.



Ref.~\cite{Moussallam:1999aq} chooses $ M = N $, which typically we take to be $ \approx 30-40 $.
Note that the method above leads to more sampling points close to the endpoints of the integration intervals.
In the elastic region, the values of $ \delta $ for which $ X $ diverges are then used as endpoints.
In the inelastic region, the function appearing in the denominator of $ \mathbf{R}^{-1} $ in Eq.~\eqref{eq:integral_equation} has zeros,
and the intervals of the numerical integration are chosen accordingly. The typical total number of integration intervals is $ \approx 20 $.

\subsection{Dealing with the polynomial ambiguity}


According to Ref.~\cite{Muskhelishvili}, there are $ n $ so-called fundamental functions $ \chi^{(i)} (s) $, $ i = 1, \ldots, n $, of lowest finite degree in the $ n $-channel coupled analysis. These solutions cannot be written as a polynomial times another solution.
Their combination with polynomial coefficients is also a solution.
The most general solution (having finite degree at infinity) is then:
%
\begin{equation}
	\sum^n_{i=1} P_i (s)\, \chi^{(i)} (s)\, ,
\end{equation}
where $ P_i (s) $ are polynomials of $ s $, and the dimension of $ \chi^{(i)} $ is $ n $.
For instance, in the two-channel coupled analysis, $ \chi^{(i)} $ are vectors of dimension two.

\begin{figure}[t]
	\centering
    \includegraphics[scale=0.38]{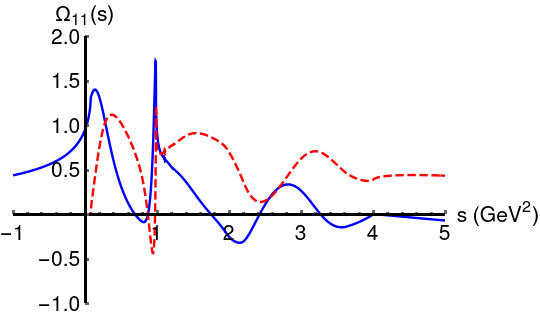} \hspace{2mm}
    \includegraphics[scale=0.38]{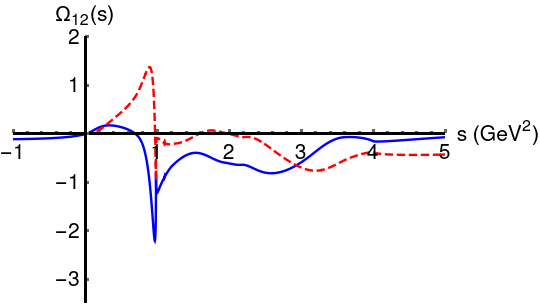} \\
    \vspace{3mm}
    \includegraphics[scale=0.38]{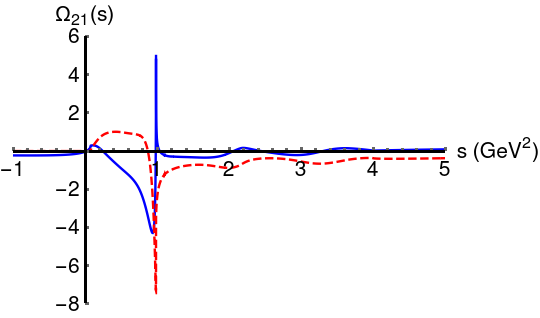} \hspace{2mm}
    \includegraphics[scale=0.38]{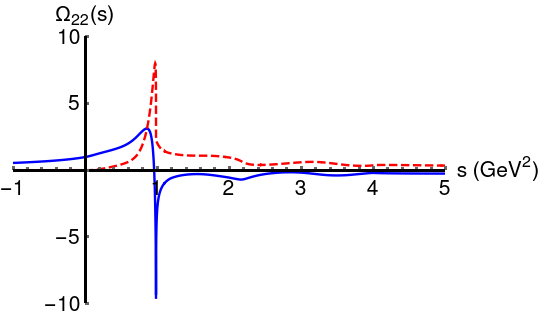}
	\caption{Set of Omn\`{e}s solutions for the reference case of Tab.~\ref{tab:num_sols_Omnes}. Real parts are shown in solid blue, while imaginary parts are shown in dashed red.}
    \label{fig:Omnes_matrix}
\end{figure}

Following the discussions of Sec.~\ref{sec:DRs} and App.~\ref{app:num_sol_DRs_method}, we generate the fundamental solutions in the latter two-channel coupled case numerically, satisfying the following condition at the subtraction point $ s_0 < 4 M^2$:
\begin{equation}
    \left(\chi^{(1)} (s_0) \, \otimes \, \chi^{(2)} (s_0)\right)\, =\; \Omega^{(0)} (s_0)\; = \,\left(\mathcal{N}^{(1)} (s_0) \, \otimes \, \mathcal{N}^{(2)} (s_0)\right) .
\end{equation}
The numerical solutions $ \mathcal{N}^{(i)}(s) $ are polynomials of degree one times the fundamental solutions $ \chi^{(i)}(s) $, as it turns out that we find numerical solutions going asymptotically to non-vanishing constants, and that the indices $ x_1 = x_2 = -1 $, see Sec.~\ref{sec:sols_coupled_channel_DRs}.
To get rid of the unknown polynomials,
we also require that another condition is satisfied at a different point $ s_1 $ (in practice, $ s_1 < s_0 $):
%
\begin{equation}
    \left(\mathcal{N}^{(1)} (s_1) \, \otimes \, \mathcal{N}^{(2)} (s_1)\right)\, =\;
    \begin{pmatrix}
    a_1 & a_3 \\
    a_2 & a_4 \\
    \end{pmatrix}\, .
\end{equation}
The values of $ a_{1,2,3,4} $, which are real, are then adjusted in order to build the matrix $ \left(\chi^{(1)} (s) \, \otimes \, \chi^{(2)} (s)\right) $ that satisfies the condition valid for the determinant, Eq.~\eqref{eq:det_equation}, for which an explicit analytical expression is known. This procedure then leads to the sought system of fundamental solutions $ \chi^{(i)} $.
They are given at $ M^2_D $ for various sets of inputs in Tab.~\ref{tab:num_sols_Omnes}. The system of fundamental solutions is shown for the reference solution in Fig.~\ref{fig:Omnes_matrix}.
(As a cross-check, with the inputs used in \cite{TarrusCastella:2021pld,TarrusCastella}, we have reproduced their Omn\`{e}s solution.)

\vspace{3mm}
We reproduce from Ref.~\cite{Muskhelishvili} the following properties of fundamental solutions that are used as checks of the previous algorithm:

\vspace{2mm}
\noindent \textbf{PROPERTY 1$ ^\text{o} $:} The determinant
\begin{equation}
	\Delta (z) = \text{det}\, || \chi^{(\beta)}_\alpha (z) || \qquad (\alpha, \beta = 1 , \ldots , n)
\end{equation}
does not vanish anywhere in the finite part of the plane.

\vspace{2mm}
\noindent \textbf{PROPERTY 2$ ^\text{o} $:} Let $x_\beta$ be the degree of the solution $ \chi^{(\beta)} (z) $ at infinity;
if one defines
\begin{equation}
	\chi^{(\beta), 0} (z) = z^{-x_\beta} \chi^{(\beta)} (z) \qquad (\beta = 1, 2, \ldots, n) ,
\end{equation}
then the determinant
\begin{equation}
	\Delta^0 (z) = \text{det}\, || \chi^{0}_\alpha (z) ||
\end{equation}
has a finite non-zero value at infinity.

Crucially, by definition any $ n $ solutions of the homogeneous Hilbert problem of Eq.~\eqref{eq:simple_DR} (where $ \mathcal{S} $ satisfies the H\"{o}lder condition ensuring it does not grow too fast with the energy \cite{Muskhelishvili}, and its determinant does not vanish, see Eq.~\eqref{eq:det_equation}),
possessing properties 1$ ^\text{o} $ and 2$ ^\text{o} $, is a fundamental system of solutions of this problem.


\vspace{3mm}
This latter step of getting rid of polynomial ambiguities has in practice been executed in \texttt{Mathematica} \cite{Mathematica}. The numerical code implemented in \texttt{Python} together with a \texttt{Mathematica} notebook containing an example will later be released in \texttt{Zenodo}.

\section{Explicit solution of the DRs close to the elastic regime}\label{app:analytic_solution}


It would be certainly important to achieve a full explicit analytical equation, instead of relying on a numerical method as described in the previous section, in order to get a higher understanding of the behaviour of the Omn\`{e}s solution given the required phase-shifts and inelasticities as inputs.
Hereafter, we discuss an explicit analytical expression for the amplitudes of the two-coupled channel problem valid close to the elastic limit.
We write Eq.~\eqref{eq:main_discontinuity_eq} as $ A = S_S \, A^\ast $.
This equation can be used to solve for the phases of the individual elements $ A_{\pi\pi}, A_{KK} $ of $A \equiv ( A_{\pi\pi}, A_{KK} )^T$ as a function of the ratio of their magnitudes:

\begin{eqnarray}
&& \cos{(\arg A_{\pi\pi}(s)-\delta_1 (s))}=\sqrt{\frac{(1+\eta(s))^2-\lambda^{-2}_{\pi K}(s) (1-\eta(s)^2)}{4\eta(s)}} \label{eq:angles_explicit_sol_1} \,,\\
&& \cos{(\arg A_{KK}(s)-\delta_2 (s))}=\sqrt{\frac{(1+\eta(s))^2-\lambda^2_{\pi K}(s) (1-\eta(s)^2)}{4\eta(s)}} \label{eq:angles_explicit_sol_2}
\end{eqnarray}
where $ \delta_1 (s) = \delta^0_0 (s) $, $ \delta_2 (s) = \psi^0_0 (s) - \delta^0_0 (s) $, $ \eta (s) = \eta^0_0 (s) $ in the isospin-zero case, and
\begin{equation}
    \lambda_{\pi K}(s)\equiv\frac{|A_{\pi\pi}(s)|}{|A_{KK}(s)|} \,.
\end{equation}
Exploiting the general once-subtracted relation arising from analyticity:
\begin{equation}
    |A_i(s)|=|A_i(s_0)| \times \exp \left\{ \frac{s-s_0}{\pi} \dashint_{4M_{\pi}^2}^{\infty} dz \frac{\arg A_i(z)}{(z-s)(z-s_0)} \right\} \,, \;\; i = \pi\pi, KK
    \label{drwitharg}
\end{equation}
where $A_i(s_0)$ collects the zeros of $ A_i(s) $,
one obtains that the ratio of the magnitudes follows:

\begin{equation}\label{lpkequation}
    \quad\quad \lambda_{\pi K}(s) = \lambda_{\pi K}(0) \times \exp \left\{ \frac{s}{\pi} \dashint_{4M_\pi^2}^{\infty}dz\frac{\delta_1(z)-\Theta(z-4M_K^2)\delta_2(z)}{z(z-s)} \right\}
\end{equation}
\vspace{-1mm}
\begin{equation}
    \times \exp \left\{ \frac{s}{\pi} \dashint_{4M_K^2}^{\infty}dz \frac{\arccos\sqrt{\frac{(1+\eta(z))^2-\lambda^{-2}_{\pi K}(z) (1-\eta(z)^2)}{4\eta(z)}}  -\arccos\sqrt{\frac{(1+\eta(z))^2-\lambda^2_{\pi K}(z) (1-\eta(z)^2)}{4\eta(z)}} }{z(z-s)}
    \right\} \nonumber
\end{equation}
for one subtraction taken at $s_0=0$.

Solving the latter equation is obviously a highly non-trivial task.
However,
close to the elastic limit $\eta(s) \sim 1$ for all relevant values of the energy $s$, we obtain the following approximation:
\begin{equation}
    \lambda^{-1}_{\pi K}(s)-\lambda_{\pi K}(s) \simeq \phi_{el}(s)+\frac{s}{\pi}g_{el}(s) \dashint_{4M_K^2}^{\infty}dz
    \frac{\epsilon(z)\phi_{el}(z)}{z(z-s)}
    \label{phi1}
\end{equation}
after expansion in the small quantity $\epsilon(s)$

\begin{equation}
\epsilon(s) \equiv \sqrt{\frac{1-\eta(s)}{2}} \,.
\end{equation}
%
The functions $ \phi_{el}(s) $ and $ g_{el}(s) $ are known from the perfect elastic limit $\eta(s) = 1$, they depend then only on the phase-shifts $ \delta_1 (s), \delta_2 (s) $ and are given by:

\begin{eqnarray}
    && \phi_{el}(s) \equiv \lambda^{-1}_{\pi K,el}(s)-\lambda_{\pi K,el}(s) \,, \\
    && g_{el}(s) \equiv -\lambda^{-1}_{\pi K,el}(s)-\lambda_{\pi K, el}(s)
\end{eqnarray}
with $\lambda_{\pi K,el} (s)$ the ratio of the amplitudes in the fully elastic case, given by the first line of Eq.~\eqref{lpkequation}. 
Having an approximation for the ratio $\lambda_{\pi K}(s)$,
the phases of the individual amplitudes can be substituted in Eq.~\eqref{drwitharg} by the use of Eqs.~\eqref{eq:angles_explicit_sol_1} and \eqref{eq:angles_explicit_sol_2}, and $A_{\pi\pi}, A_{KK}$ can be obtained as functions of $s$.
A drawback of this approach is that the ratio $\lambda_{\pi K, el} (s)$ may get close to zero, rendering ill-defined the procedure described above, being well behaved for $ \lambda_{\pi K, el}(s) \sim 1 $.
Due to these shortcomings, we stress that such a method, which illustrates the difficulty in obtaining an explicit analytical solution, has not been employed in the present work.

\section{Decay constants and form factors}\label{app:norm_sign_convs}

We need the following hadronic matrix elements of the axial-vector (no sum over $i,j$ is implied),
\begin{equation}
\langle 0| \bar q^j\gamma^\mu\gamma_5 q^i | P^{ij}(p)\rangle 
\, =\, -\langle P^{ji}(p)| \bar q^j\gamma^\mu\gamma_5 q^i | 0\rangle
\, =\, i\, C_P^{ij}\, f_P\, p^\mu\, ,
\end{equation}
and vector,
\begin{equation}\label{eq:VectorMatrixElements}
\langle P'(p')| \bar q^j\gamma^\mu q^i | P(p)\rangle \, =\,  \widetilde{C}_{PP'}^{ij}\left[ (p+p')^\mu\, f_+^{PP'}(q^2) + (p-p')^\mu\, f_-^{PP'}(q^2)\right] ,
\end{equation}
QCD currents, where $q^\mu = p^\mu-p'^\mu$ 
and the superindices in $P^{ij}\sim q^i\bar q^j$ indicate the flavour content of the corresponding pseudoscalar meson (they are not displayed explicitly in the vector case where flavour quantum numbers can match in different ways).

In the axial-vector matrix element, the normalization of the decay constant corresponds to 
$f_\pi = \sqrt{2} F_\pi = (130.2\pm 0.8)$~MeV \cite{FlavourLatticeAveragingGroupFLAG:2021npn}.
The coefficient $C_P^{ij}$ reflects the intrinsic flavour composition of $P^{ij}$.
It is just equal to 1 for flavourful mesons, while for the flavourless states:
$$
C^{11}_{\pi^0} = - C^{22}_{\pi^0} = \frac{1}{\sqrt{2}}\, ,
\qquad\quad
C^{11}_{\eta_8} = C^{22}_{\eta_8} = -\frac{1}{2}\, C^{33}_{\eta_8} = \frac{1}{\sqrt{6}}\, ,
\qquad\quad
C^{ii}_{\eta_0}=\frac{1}{2}\, , 
$$
\begin{equation}\label{eq:Ccoefficients}
C^{11}_{\eta_{15}} = C^{22}_{\eta_{15}} = C^{33}_{\eta_{15}} =-\frac{1}{3}\, C^{44}_{\eta_{15}} = \frac{1}{\sqrt{12}}\, . 
\end{equation}
These factors are conveniently captured in the following $4\times 4$ matrix of pseudoscalar bosons \cite{Botella:1993va}:
\begin{equation}
\label{eq:phi-chpt}
\Phi\, =\,\begin{pmatrix}
\frac{\pi^0}{\sqrt{2}}+ \frac{\eta_8}{\sqrt{6}} + \frac{\eta_{15}}{\sqrt{12}} + \frac{\eta_0}{2} & \pi^+ & K^+ & \overline{D}^0
\\ 
\pi^- & -\frac{\pi^0}{\sqrt{2}}+ \frac{\eta_8}{\sqrt{6}} + \frac{\eta_{15}}{\sqrt{12}} + \frac{\eta_0}{2} & K^0 & D^-
\\
K^- & \overline{K}^0 & -\frac{2\eta_8}{\sqrt{6}} + \frac{\eta_{15}}{\sqrt{12}} + \frac{\eta_0}{2} & D^-_s
\\
D^0 &D^+ & D^+_s & -\frac{3\eta_{15}}{\sqrt{12}} + \frac{\eta_0}{2}
\end{pmatrix} ,
\end{equation}
which fixes our conventions. Under charge conjugation $\Phi\to\Phi^T$.
In the unphysical limit of vanishing quark masses, the  axial quark current has the effective hadronic representation $\bar q^j\gamma^\mu\gamma_5 q^i\dot= -f\,\partial^\mu\Phi^{ij} +\mathcal{O}(\Phi^3)$, while the vector current is given by $\bar q^j\gamma_\mu q^i\dot= -i\left(\Phi
\!\stackrel{\leftrightarrow}{\partial}_\mu\!\Phi\right)^{\! ij} + \mathcal{O}(\Phi^4)$ \cite{Pich:1995bw}. This reproduces the constant factors in Eq.~(\ref{eq:Ccoefficients}) and allows one to easily derive the appropriate Clebsh-Gordon coefficients in Eq.~(\ref{eq:VectorMatrixElements}), because the vector-current matrix element satisfies $f_+^{PP'}(0)=1$ in the massless quark limit (vector-current conservation). We only quote here those coefficients needed in our calculation:
\begin{equation}
\widetilde{C}_{D^+\pi^+}^{41} = \sqrt{2}\, \widetilde{C}_{D^0\pi^0}^{41} =
\widetilde{C}_{D^0\pi^-}^{42} = -\sqrt{2}\, \widetilde{C}_{D^+\pi^0}^{42} =
\widetilde{C}_{D^0 K^-}^{43} = \widetilde{C}_{D^+ \overline{K}^0}^{43} =
 1\, .
\end{equation}

Since 
\begin{equation}
q_\mu\; \langle P'(p')| \bar q^j\gamma^\mu q^i | P(p)\rangle \, =\,  \widetilde{C}_{PP'}^{ij}\left( M_P^2-M_{P'}^2\right)  f_0^{PP'}(q^2)\, ,
\end{equation}
the scalar form factor
\begin{equation}
f_0^{PP'}(q^2)\, = \, f_+^{PP'}(q^2) +\frac{q^2}{ M_P^2-M_{P'}^2}\, f_-^{PP'}(q^2)
\end{equation}
plays an important role in the bare decay amplitudes.

For the evaluation of the penguin contribution ($Q_6$), we also need the scalar and pseudoscalar matrix elements, which can be easily obtained by applying the QCD equations of motion:
\begin{eqnarray}
\langle 0| \bar q^j \gamma_5 q^i | P^{ij}(p)\rangle &\!\!\! =&\!\!\!  
-i\,\frac{\langle 0| \partial_\mu (\bar q^j\gamma^\mu\gamma_5 q^i) | P^{ij}(p)\rangle}{m_i + m_j}
\, =\, -i\, C_P^{ij}\,\frac{ f_P\, M_P^2}{m_i + m_j}\, ,
\\
\langle P(p')| \bar q^j q^i | P(p)\rangle &\!\!\! =&\!\!\! 
i\,\frac{\langle P(p')| \partial_\mu (\bar q^j\gamma^\mu q^i) | P(p)\rangle}{m_i-m_j} \, =\, \frac{ \widetilde{C}_P^{ij}}{m_i-m_j}\,
\left( M^2_P-M^2_{P'}\right) f_0^{PP'}(q^2)\, .
\nonumber\end{eqnarray}

For equal quark masses the needed two-Goldstone matrix elements of the  light-quark scalar currents,
\begin{eqnarray}
&& \hskip -1cm
\langle\pi^i |\bar u u +\bar d d | \pi^j\rangle \, = \, \delta^{ij}\; F^\pi_S(t)\, ,
\nonumber\\
\langle K^+ |\bar u u | K^+\rangle &\!\! = &\!\! 
\langle K^+ |\bar u d | K^0\rangle \, =\, 
\langle K^0 |\bar d d | K^0\rangle \, =\,
F^K_S(t)\, ,
\end{eqnarray}
can be determined at low momentum transfer with $\chi$PT \cite{Gasser:1984gg,Pich:1995bw}.
At $\mathcal{O}(p^4)$ and keeping only the leading contributions at large-$N_C$, one gets: 
\begin{eqnarray}\label{eq:leading_Nc_scalar_FFs}
F^\pi_S(t) &\!\! = &\!\! \frac{M_\pi^2}{\hat m}\,\left\{ 1 + 
\frac{16}{f_\pi^2}\, (2  L_8-L_5)\, M_\pi^2 + \frac{8 L_5}{f_\pi^2}\, t \right\}
\,\equiv\, \frac{M_\pi^2}{\hat m}\; \widetilde F^\pi_S(t)\, ,
\nonumber\\
F^K_S(t)&\!\! = &\!\!  \frac{M_K^2}{m_s+\hat{m}}\,\left\{ 1 + 
\frac{16}{f_K^2}\, (2  L_8-L_5)\, M_K^2 + \frac{8 L_5}{f_K^2}\, t
\right\} 
\,\equiv\, \frac{M_K^2}{m_s+\hat{m}}\; \widetilde F^K_S(t)\, ,
\end{eqnarray}
with $\hat m = m_u = m_d$.
For the chiral low-energy constants we will adopt the values $L_5^r(M_\rho) = (1.20\pm 0.10) \times 10^{-3}$ and 
$2  L_8-L_5 = -(0.15\pm 0.20) \times 10^{-3}$
\cite{Cirigliano:2019cpi}.

\section{Bare decay amplitudes}
\label{app:matrix_elements}

The hadronic matrix elements of the four-quark operators in Eq.~(\ref{eq:operator_list}) are non-perturbative quantities, sensitive to the involved infrared properties of the strong interaction. However, they can be easily evaluated in the limit of a large number of QCD colours, because the product of two colour-singlet quark currents factorizes at the hadron level into two current matrix elements \cite{Buras:1985xv,Gisbert:2017vvj}:
\begin{equation}
\langle J\cdot J\rangle \, =\, \langle J \rangle \,  \langle J \rangle\;\left\{ 1 + \mathcal{O}\left(\frac{1}{N_C}\right)\right\} .
\end{equation}
For instance, when $N_C\to\infty$,
%
\begin{eqnarray}
\lefteqn{\langle \pi^-\pi^+ | (\bar d c)_{V-A} (\bar u d)_{V-A}   | D^0\rangle \, =\,
\langle\pi^- | (\bar d c)_{V-A} | D^0\rangle\;
\langle \pi^+| (\bar u d)_{V-A}   | 0\rangle} &&
\nonumber\\ &&\hskip 1cm =\,
-\langle\pi^- | \bar d \gamma_\mu c | D^0\rangle\;
\langle \pi^+| \bar u\gamma^\mu\gamma_5 d   | 0\rangle
\, =\,
i f_\pi\left(M_D^2-M_\pi^2\right) f_0^{D\pi}(M_\pi^2)
\, ,
\end{eqnarray}
while the penguin $Q_6$ operator gives 
\begin{eqnarray}\label{eq:Q6MatrixEl}
\lefteqn{-2\,\sum_q \, \langle \pi^-\pi^+ | (\bar q c)_{S-P} (\bar u q)_{S+P} | D^0\rangle } &&
\nonumber\\ &&\hskip 1.5cm =\, 
2\, \langle 0 | \bar u\gamma_5 c | D^0\rangle\;
\langle \pi^-\pi^+| \bar u u   | 0\rangle
-2\, \langle\pi^- | \bar d c | D^0\rangle\; \langle \pi^+| \bar u\gamma_5 d   | 0\rangle
\nonumber\\ &&\hskip 1.5cm =\,
-2 i\, \frac{M_\pi^2}{2\hat m}\left[\frac{f_D M_D^2}{m_c+\hat{m}}\,\widetilde F_S^\pi(M_D^2) + \frac{f_\pi \left( M_D^2-M_\pi^2\right)}{m_c-\hat m}\, f_0^{D\pi}(M_\pi^2)
\right] .
\end{eqnarray}

Using the matrix elements of the QCD currents given in appendix~\ref{app:norm_sign_convs}, one can then determine all bare decay amplitudes in the large-$N_C$ limit:
%
\begin{eqnarray}\label{eq:BareAmplitudes}
T_{D^0\to\pi^-\pi^+}^{(B)} &\!\! = &\!\! \frac{G_F}{\sqrt{2}}\, f_\pi \left(M_D^2-M_\pi^2\right) f_0^{D\pi}(M_\pi^2) \left[\lambda_d\, C_1 - \lambda_b\, (C_4 - C_6\,\delta_6^\pi)\right] ,
\nonumber\\
T_{D^0\to\pi^0\pi^0}^{(B)} &\!\! = &\!\! -\frac{G_F}{\sqrt{2}}\, f_\pi \left(M_D^2-M_\pi^2\right) f_0^{D\pi}(M_\pi^2) \left[\lambda_d\, C_2 + \lambda_b\, (C_4- C_6\,\delta_6^\pi)\right] ,
\nonumber\\
T_{D^+\to\pi^0\pi^+}^{(B)} &\!\! = &\!\! -\frac{G_F}{\sqrt{2}}\,\frac{f_\pi}{\sqrt{2}} \left(M_D^2-M_\pi^2\right) f_0^{D\pi}(M_\pi^2) \,\lambda_d\, (C_1 + C_2)\, ,
\nonumber\\
T_{D^0\to K^- K^+}^{(B)} &\!\! = &\!\! \frac{G_F}{\sqrt{2}}\, f_K \left(M_D^2-M_K^2\right) f_0^{DK}(M_K^2) \left[\lambda_s\, C_1 - \lambda_b\, (C_4- C_6\,\delta_6^K)\right] ,
\nonumber\\
T_{D^0\to \overline{K}^0 K^0}^{(B)} &\!\! = &\!\!  0\, ,
\nonumber\\
T_{D^+\to \overline{K}^0 K^+}^{(B)} &\!\! = &\!\! \frac{G_F}{\sqrt{2}}\,
 f_K \left(M_D^2-M_K^2\right) f_0^{DK}(M_K^2) \left[\lambda_s\, C_1 - \lambda_b\, (C_4- C_6\,\delta_6^K)\right] ,
\end{eqnarray}
where
\begin{eqnarray}
\delta_6^\pi &\!\! = &\!\! \dfrac{2}{m_c-\hat m} \,\dfrac{M_\pi^2}{2\hat m}\left\{1 + \dfrac{f_D M_D^2}{f_\pi (M_D^2-M_\pi^2)}\, \dfrac{m_c-\hat m}{m_c+\hat m}\,\dfrac{\widetilde F_S^\pi(M_D^2)}{f_0^{D\pi}(M_\pi^2)}\right\} ,
\nonumber\\
\delta_6^K &\!\! = &\!\! \dfrac{2}{m_c- m_s}\,\dfrac{M_K^2}{m_s+\hat m} \left\{1 + \dfrac{f_D M_D^2}{f_K (M_D^2-M_K^2)}\, \dfrac{m_c- m_s}{m_c+ \hat m}\,\dfrac{\widetilde F_S^K(M_D^2)}{f_0^{DK}(M_K^2)}\right\} .
\end{eqnarray}

The conservation of the vector current guarantees that annihilation topologies give zero contribution, except for $Q_6$ which has a scalar-pseudoscalar structure. The matrix elements of $Q_3$ and $Q_5$ are also identically zero at $N_C\to\infty$ because $\sum_i \bar q_i\gamma_\mu\gamma_5 q_i$ only couples to isosinglet states.

The bare decay amplitudes involve the hadronic parameters $f_\pi$, $f_K$, $f_0^{D\pi}(M_\pi^2)$ and $f_0^{DK}(M_K^2)$, which we take from lattice calculations. These ``physical'' inputs include  higher-order contributions in the $1/N_C$ expansion, dressing in this way the current matrix elements beyond the large-$N_C$ approximation. These additional corrections are totally independent of the rescattering dynamics incorporated in $\Omega^{(I)}(s)$.

A subtlety arises with the annihilation contribution to the matrix elements of the operator $Q_6$, given for the $\pi^+\pi^-$ case by the first term in Eq.~(\ref{eq:Q6MatrixEl}). This introduces the parameters $F_S^\pi(M_D^2)$ and $F_S^K(M_D^2)$ at $N_C\to\infty$, which are subjected to a large uncertainty. Their physical values at $N_C=3$ are fully entangled with the rescattering dynamics of the final pair of pseudoscalars.\footnote{The calculation of these scalar form factors is interesting on its own. We defer to a forthcoming publication a detailed analysis of our predicted form factors and their comparison with previous calculations.} Using crossing symmetry, we input the $\chi$PT predictions in Eq.~(\ref{eq:leading_Nc_scalar_FFs}) at the subtraction point $s_0$ and let our calculated rescattering matrix to generate the physical form factors at $s=M_D^2$.\footnote{Owing to the small value of $\mathrm{Re}\{\lambda_b\}$, the $D^0$ decay branching ratios are not sensitive to the penguin operators and, therefore, the scalar form factors do not contaminate the specification of $\Omega^{(0)}(s)$.}

The global quark-mass factors in $\delta_6^{\pi,K}$ introduce an explicit dependence on the short-distance renormalization scale that exactly cancels the corresponding dependence of the Wilson coefficient $C_6(\mu^2)$, in the large-$N_C$ limit. $Q_6$ is in fact the only four-quark operator with a non-zero anomalous dimension in the limit $N_C\to\infty$ \cite{Bardeen:1986uz}. In order to keep all short-distance logarithmic contributions, the Wilson coefficients are fully computed at NLO, without any $1/N_C$ expansion. Therefore, a subleading dependence on $\mu$ remains.


\subsection{Isospin decomposition}
\label{app:isospin_decomposition}

Bose symmetry only allows an $S$-wave $2\pi$ state to have $I=0$ and 2. In terms of isospin states $|I, I_3\rangle$ the $2\pi$ final states with definite charges are decomposed as:\footnote{We adopt the usual isospin convention with quark multiplets $(u,d)$ and $(-\bar d,\bar u)$, and meson multiplets $(-\pi^+,\pi^0,\pi^-)$, $(K^+,K^0)$, $(-\overline{K}^0, K^-)$, $(\overline{D}^0, D^-)$ and $(-D^+,D^0)$, which is consistent with the matrix realization in Eq.~(\ref{eq:phi-chpt}).}
\begin{eqnarray}
|\pi^0\pi^0\rangle & = & \sqrt{\frac{2}{3}}\; |2,0\rangle - \frac{1}{\sqrt{3}}\; |0,0\rangle\, ,
\nonumber\\
\frac{1}{\sqrt{2}}\, |\pi^+\pi^- + \pi^-\pi^+\rangle & = &
- \frac{1}{\sqrt{3}}\; |2,0\rangle - \sqrt{\frac{2}{3}}\; |0,0\rangle\, ,
\nonumber\\
\frac{1}{\sqrt{2}}\, |\pi^+\pi^0 + \pi^0\pi^+ \rangle & = &
- |2,1\rangle\, .
\end{eqnarray}
Therefore,\footnote{$\langle I^f I_3^f|O_{I I_3}|I^i I_3^i\rangle = 
\langle  I I_3 I^i I_3^i| I I^i I^f I_3^f\rangle\; \langle I^f||O_I|| I^i\rangle$.
The factor $1/\sqrt{2}$ in front of the $\pi^-\pi^+$ and $\pi^0\pi^+$ amplitudes reabsorbs the phase-space factor for identical particles, so that one recovers the usual normalization of distinguishable particles adopted in the dynamical calculations.}
%
\begin{eqnarray}
A[D^0\to\pi^0\pi^0] &\! \! = &\! \! - \frac{1}{\sqrt{6}}\; T^0_{\pi\pi} + \frac{1}{\sqrt{3}}\; T^2_{\pi\pi} \, ,
\nonumber\\
A[D^0\to\pi^-\pi^+]&\! \!\equiv &\! \! \frac{1}{\sqrt{2}}\, A[D^0\to\frac{1}{\sqrt{2}}\, (\pi^+\pi^-+\pi^-\pi^+)] \, = \,  -\frac{1}{\sqrt{6}}\; T^0_{\pi\pi} -  \frac{1}{2\sqrt{3}}\; T^2_{\pi\pi} \, ,
\nonumber\\
A[D^+\to\pi^0\pi^+] &\! \!\equiv &\! \! \frac{1}{\sqrt{2}}\, A[D^+\to\frac{1}{\sqrt{2}}\, (\pi^+\pi^0+\pi^0\pi^+)] \, = \,  \frac{\sqrt{3}}{2\sqrt{2}}\; T^2_{\pi\pi} \, .
\end{eqnarray}
%

%

The $K\overline{K}$ system can have $I=0$ and $I=1$:
\begin{eqnarray}
|K^- K^+\rangle &=&\frac{1}{\sqrt{2}}\; |1,0\rangle - \frac{1}{\sqrt{2}}\;|0,0\rangle\, ,
\nonumber\\
|\overline{K}^0 K^0\rangle &=& -\frac{1}{\sqrt{2}}\; |1,0\rangle - \frac{1}{\sqrt{2}}\; |0,0\rangle\, ,
\nonumber\\
|\overline{K}^0 K^+\rangle &=& - |1, 1\rangle \, .
\end{eqnarray}
This implies
%
\begin{eqnarray}
A(D^0\to K^- K^+) &=& \frac{1}{2} \left( T^{11}_{KK} +  T^{13}_{KK} - T^0_{KK}\right) ,
\nonumber\\
A(D^0\to \overline{K}^0 K^0) &=& \frac{1}{2}\, \left( -T^{11}_{KK} -  T^{13}_{KK} - T^0_{KK}\right) ,
\nonumber\\
A(D^+\to \overline{K}^0 K^+) &=& T^{11}_{KK} - \frac{1}{2}\, T^{13}_{KK}\, .
\end{eqnarray}
Here, $T^{11}_{KK}$ and $T^{13}_{KK}$ denote the reduced amplitudes
$\langle 1||O_{1/2}||\frac{1}{2}\rangle$ and $\langle 1||O_{3/2}||\frac{1}{2}\rangle$, respectively.

In the large-$N_C$ limit, we get from Eq.~(\ref{eq:BareAmplitudes}):
%
\begin{eqnarray}\label{eq:BareIsospinAmplitudes}
T^{0\, (B)}_{\pi\pi} &\!\! = &\!\!
-\frac{G_F}{\sqrt{2}}\, \sqrt{\frac{2}{3}}\, f_\pi \left(M_D^2-M_\pi^2\right) f_0^{D\pi}(M_\pi^2) \left[\lambda_d\, (2 C_1-C_2) - 3\lambda_b\, (C_4 - C_6\,\delta_6^\pi)\right] ,
\nonumber\\
T^{2\, (B)}_{\pi\pi} &\!\! = &\!\!
-\frac{G_F}{\sqrt{2}}\,\frac{2f_\pi}{\sqrt{3}} \left(M_D^2-M_\pi^2\right) f_0^{D\pi}(M_\pi^2) \,\lambda_d\, (C_1 + C_2)\, ,
\nonumber\\
- T^{0\, (B)}_{KK} &\!\! = &\!\!
T^{11\, (B)}_{KK} \, =\, 
\frac{G_F}{\sqrt{2}}\,
 f_K \left(M_D^2-M_K^2\right) f_0^{DK}(M_K^2) \left[\lambda_s\, C_1 - \lambda_b\, (C_4- C_6\,\delta_6^K)\right] ,
\nonumber\\
T^{13\, (B)}_{KK} &\!\! = &\!\! 0 \, .
\end{eqnarray}
%


\bibliography{mybib}{}
\bibliographystyle{unsrturl}

\end{document}